\documentclass[namedreferences]{SolarPhysics}
\usepackage[optionalrh]{spr-sola-addons} 
\usepackage{graphicx}                    
\usepackage{color}                       
\usepackage{url}                         
\usepackage{verbatim}

\def\apjl{Astrophys. J.}%
\def\apjs{ApJS}%
\def\ao{Appl.~Opt.}%
\def\aap{Astron. Astrophys.}%
\usepackage{multirow}

\def\arcsec{\hbox{$^{\prime\prime}$}}

\newcommand{\arcs}{\ensuremath{^{\prime\prime}}}

\newcommand{\dl}{~{\mathrm d} l}
\newcommand{\dA}{~{\mathrm d} A}

\def\ion[#1 #2]{#1\,{\sc #2}}
\def\ergs[#1]{#1 {ergs}~{cm$^{-2}$}\,{s$^{-1}$}\,{sr$^{-1}$}}
\def\dens[#1]{10$^{#1}$\hskip 1.5pt{cm$^{-3}$}}
\def\densr[#1 #2]{10$^{#1}$\hskip 1pt{---}\hskip .5pt{10$^{#2}$}\hskip 1.5pt{cm$^{-3}$}}
\def\fl[#1 #2]{{#1}$\pm${#2}}
\def\orb[#1 #2]{{$#1^{#2}$}}
\def\ls[#1 #2]{{$^{#1}${#2}}}
\def\tm[#1 #2 #3]{{$^{#1}${#2}$_{#3}$}}

\begin{document}
\begin{article}

\begin{opening}

\title{The Structure and Dynamics of the Upper Chromosphere
  and Lower Transition Region as Revealed by the Subarcsecond VAULT
  Observations} 
\author{A.~\surname{Vourlidas}$^{1}$\sep
B.~\surname{Sanchez Andrade-Nu\~{n}o}$^{1,2}$\sep
E.~\surname{Landi}$^{1}$\sep
S.~\surname{Patsourakos}$^{2}$\sep
L.~\surname{Teriaca}$^{3}$\sep
U.~\surname{Sch\"{u}hle}$^{3}$\sep
C.M.~\surname{Korendyke}$^{1}$\sep
I.~\surname{Nestoras}$^{4}$
       }
\institute{$^{1}$ Space Science Division, Naval Research
Laboratory, 4555 Overlook Ave, SW, Washington, D.C., USA
\email{vourlidas@nrl.navy.mil}}
\institute{$^{2}$ George Mason University, 4400 University Dr, Fairfax, VA, USA
\email{bsanchez@ssd5.nrl.navy.mil}}
\institute{$^{3}$ MPI for Solar System Research, 37191 Katlenburg-Lindau, Germany
\email{teriaca@mps.mpg.de, schuele@mps.mpg.de}}
\institute{$^{4}$ Max-Plank-Institut f\"{u}r Radioastronomie, Auf dem
  H\"{u}gel 69, 53121 Bonn, Germany}  

\begin{abstract}
  The Very high Angular resolution ULtraviolet Telescope (VAULT) is a
  sounding rocket payload built to study the crucial interface between
  the solar chromosphere and the corona by observing the strongest
  line in the solar spectrum, the Ly$\alpha$ line at 1216\AA. In two
  flights, VAULT succeeded in obtaining the first ever sub-arcsecond
  ($0.5\arcsec$) images of this region with high sensitivity and
  cadence. Detailed analyses of those observations have contributed
  significantly to new ideas about the nature of the transition
  region. Here, we present a broad overview of the Ly$\alpha$
  atmosphere as revealed by the VAULT observations, and bring together
  past results and new analyses from the second VAULT flight to create
  a synthesis of our current knowledge of the high-resolution
  Ly$\alpha$ Sun. We hope that this work will serve as a good
  reference for the design of upcoming Ly$\alpha$ telescopes and
  observing plans.
\end{abstract}

\keywords{line: Hydrogen Ly alpha ---- atomic data ---- Sun: corona ---- Sun:
  UV radiation ---- Sun: transition region}

\end{opening}

%
 \section{Introduction \label{s:Intro}} 

 The structure of the solar atmosphere as a function of
 temperature has been a 'thorny' issue of solar physics research for
 decades. As the density decreases, the temperature, instead of
 decreasing, abruptly increases from $\sim 10^4$ K to a million K
 within a thousand km. It is known since the first solar imaging space
 missions that this so-called temperature transition region (TR)
 between the chromosphere and the corona, is also where the morphology
 of the atmospheric structures changes strongly.  At the base of the
 atmosphere, the photosphere consists of small scale convective
 granules interlaced with occasional smaller intergranular lanes
 concentrating strong magnetic flux elements($|B|\le1kG$,
 e.g. \opencite{2004Natur.430..326T}).  The chromosphere (T$\leq
 10^4$ K for the discussion here) consists of a very rugged,
 inhomogeneous, and very filamentary layer blanketing the photosphere.

 Beginning at the chromosphere, the geometry of the individual
 structures is increasingly dominated by the local magnetic field. At
 the lower transition region (T$\leq 2\times 10^5$ K), the structures
 strongly reflect the morphology of the underlying supergranular
 network. As the magnetic pressure overtakes the gas pressure leading
 to the low beta corona, the percentage of emission in filamentary
 loops steadily increases until the network completely disappears at
 temperatures above $10^6$ K. It may seem that a straightforward
 interplay between heating and morphology takes place in the transition
 region but this is not the case.

 The traditional picture of the transition region as the interface
 between the footpoints of large-scale structures and their coronal
 tops has been contradicted by the weakness of its observed emission
 \cite{2004ApJ...611..537L}. While the emission in the upper TR
 ($T>2\times 10^5$K) can be understood in terms of heat conduction
 from the corona along magnetic field lines, the lower TR ($T<2\times
 10^5$ K) cannot. Instead, this plasma forms a completely separate
 component of the solar atmosphere
 \cite{1983ApJ...275..367F,1987ApJ...320..426F}. This component could
 consists of small “cool” loops
 \cite{1986ApJ...301..440A,1986SoPh..105...35D,2000SSRv...93..411F,2001A&A...374.1108P}
 that are best seen in the Quiet Sun and that probably correspond to
 the upper reaches of the mixed polarity magnetic carpet
 \cite{1997ApJ...487..424S}. \inlinecite{2009ApJ...693.1474F} showed
 that the Differential Emission Measure (DEM) of the TR has the same
 shape everywhere (coronal holes, Quiet Sun, active regions) while
 coronal DEM of the very same regions are very different. Why and how
 are transition region loops different from higher arching coronal
 loops?  Are they also comprised of unresolved strands? Are they
 heated in a fundamentally different way? Recently, Judge (2008)
 proposed a radically different view of the transition region
 emission, suggesting that it might result from cross-field diffusion
 of plasma from very fine cool threads extending into the corona
 (e.g. spicules), and its subsequent ionization. Cool threads
 gradually expand in thickness as the ionizing front expands across
 the field lines and emits at TR temperatures, and provide images of
 the transition region similar to those observed by the SUMER
 \cite{1995SoPh..162..189W} spectrometer aboard SOHO.

 \textsl{Hinode\/} observations revealed a dramatically new picture of
 the solar chromosphere and demonstrated its potential importance for
 the dynamics, energy and mass supply of the transition region and
 corona. High temporal ($\simeq$5s) and spatial ($\simeq$0.2\arcsec)
 Hinode/SOT observations have shown that the chromosphere is much more
 structured and dynamic than previously believed.  SOT has revealed a
 chromosphere hosting a wealth of wave and oscillatory phenomena
 manifested as longitudinal and transverse motions within structures
 at the resolution limit \cite{2007PASJ...59S.655D,2008A&A...482L...9O,2007Sci...318.1577O}. Even a fraction of the
 inferred wave energy flux could account for the coronal energy losses
 if it reached the corona.  SOT also showed that a significant
 fraction of observed spicules ('type II'), known for decades to
 dominate the chromospheric landscape, disappear very rapidly (De
 Pointieu et al. 2007b). This was interpreted as a signature of the
 plasma heating up to transition region and coronal temperatures; the
 mass contained in these disappearing spicules is sufficient to
 account for the mass present in the corona.

 Capturing the fine spatial scales and rapid temporal evolution of the
 chromosphere and transition region plasmas represents a considerable
 observational and technical challenge. Nonetheless, recent significant
 improvements on instrumentation and image processing has been
 achieved both from ground (e.g., \opencite{2006A&A...454.1011P},
 \opencite{2008SoPh..251..533R}, \opencite{2007ASPC..368...65D}) or
 spaceborne instruments (e.g. \opencite{2007PASJ...59S.655D}),
 reaching in all cases spatial resolution under 1\arcsec\ for plasmas
 at chromospheric regimes. Reaching these resolution on the TR
 involves the use of strong UV lines, accessible only above the Earth's
 atmosphere.

 The Very high Angular resolution ULtraviolet Telescope (VAULT,
 \opencite{2001SoPh..200...63K}), a sounding rocket payload, is the
 only instrument that has observed this critically important layer of
 the solar atmosphere at such high resolution.  VAULT is specifically
 designed to obtain high spectral purity, zero dispersion
 spectroheliograms in the Lyman-$\alpha$ (1216 \AA) resonance line of
 hydrogen.  This emission line emanates from plasmas at 8000 to 30000K
 \cite{1986A&A...154..154G}.  The Ly$\alpha$ radiation directly maps
 the dominant energy loss from plasmas at these temperatures which
 correspond to the lower TR \cite{1988ApJ...329..464F}.  This
 instrument is the latest in a long and distinguished line of solar
 optical instruments obtaining observations in the Ly$\alpha$ emission
 line \cite{1972ApJ...176..239P,1974ApJ...187..369P,1975BAAS....7..432B,1980ApJ...237L..47B}. The VAULT observations
 are the highest quality UV observations of the solar atmosphere ever
 obtained and are a considerable improvement over previous
 instruments.  Each rocket flight obtained observations with
 observable structures of $<0.5\arcsec$ spatial scale, exposure times of
 1 second with a 17 second cadence and a 355\arcsec$\times$235\arcsec\
 instantaneous field of view (FoV).

 The VAULT data and, more recently, the \textsl{Hinode\/}/SOT
 observations have invigorated the debate about the nature of the
 solar Transition Region. Not surpsingly, Ly$\alpha$ telescopes are
 planned for the upcoming \textsl{Solar Orbiter\/} mission, and,
 possibly, the proposed \textsl{Solar-C} mission. It is therefore, an
 appropriated time for a review of the VAULT observations. We believe
 that as a trailblazer project in the exploration of the upper
 chromosphere-corona interface, the VAULT experiences will be a useful
 reference for the instrument design and science operations for those
 missions. We also take this opportunity to present the final
 calibration of the data and introduce the project website where all
 the data are publicly available.  This paper presents a detailed
 examination of the Ly$\alpha$ structures near the base of the solar
 corona obtained during the second flight of the payload (hereafter,
 VAULT-II). We are specifically concerned with those plasmas whose
 temperatures lie between 8\,000 and 30\,000K, ranging roughly from
 $\sim$2\,000 km to $\sim$60\,000~km above the photosphere.

The paper is organized as follows. Section \ref{s:calib} describes the
latest instrument calibration and the observations from the second VAULT
flight. Section \ref{s:ly} summarizes the importance of the Ly$\alpha$
in the frame of Coronal and TR models.  Section \ref{s:inten}
discusses the sources of Ly$\alpha$ emission as determined in the
VAULT images. Sections \ref{s:prom},
\ref{s:qs}, \ref{s:spicules} focus on, respectively, prominences,
Quiet Sun, and spicules. We discuss
our findings and conclude in Sections \ref{s:dis}-\ref{s:con}.

\begin{figure}[p]
 \centerline{\includegraphics[width=\textheight, angle=-90]{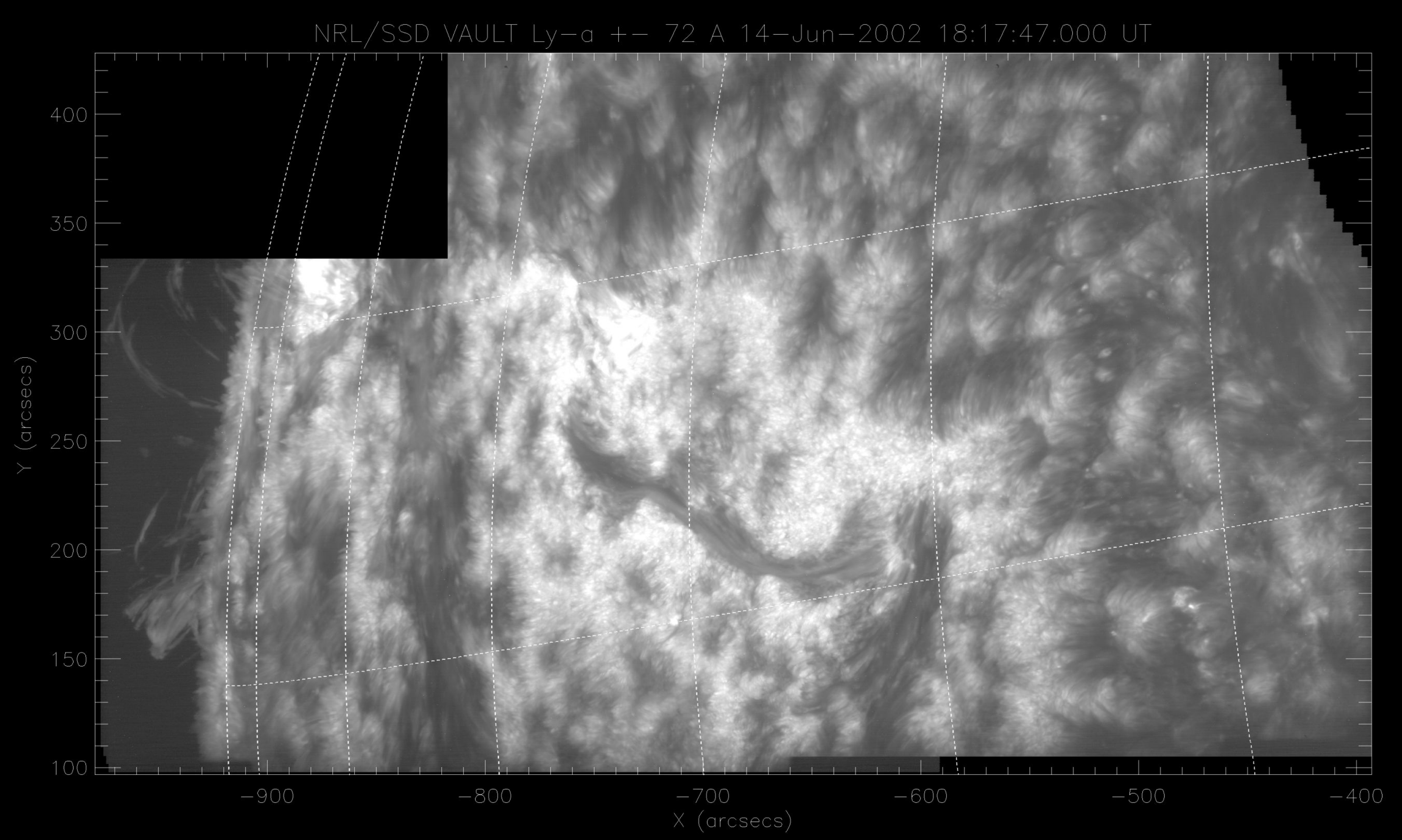}}
 \caption{The total solar field of view observed during the second VAULT
   flight. The image is a composite of all VAULT-II observations after
   dark current subtraction and flatfielding. It
   covers a $\sim600\times450$ arcsecs area, with $0.12\arcsec$ pixel
   size. Solar North is to the right and Solar East at the top of the
   image. The image is plotted with histogram equalization of the intensities.}\label{fig:composite}
 \end{figure}

\section{Data Analysis and Observations\label{s:calib}}
VAULT has been successfully launched twice (May, 7, 1999 and June, 14,
2002). Using the experience from the first flight
\cite{2001SoPh..200...63K}, the instrument performance during the
second flight was improved by
using a higher transmission Ly$\alpha$ filter (higher
throughput) and better filtering of the power converter output (lower
noise/higher quality data). So we concentrate on the VAULT-II images
for the remainder of the paper.

VAULT-II was flown on June 14, 2002 from White Sands Missile Range
onboard a Black Brant sounding rocket. The observations took place
around the apogee of the parabolic trajectory while the rocket was
above 100 km. This minimum altitude was chosen to minimize absorption
effects from the geocorona \cite{1977JGR....82.1481P}. The duration of science
operations was $363$ sec and the rocket peaked at an altitude of 182
miles (294 km). The entire flight, from launch to recovery, lasted 15
minutes. 

\begin{figure}
 \centerline{\includegraphics[width=0.50\textwidth]{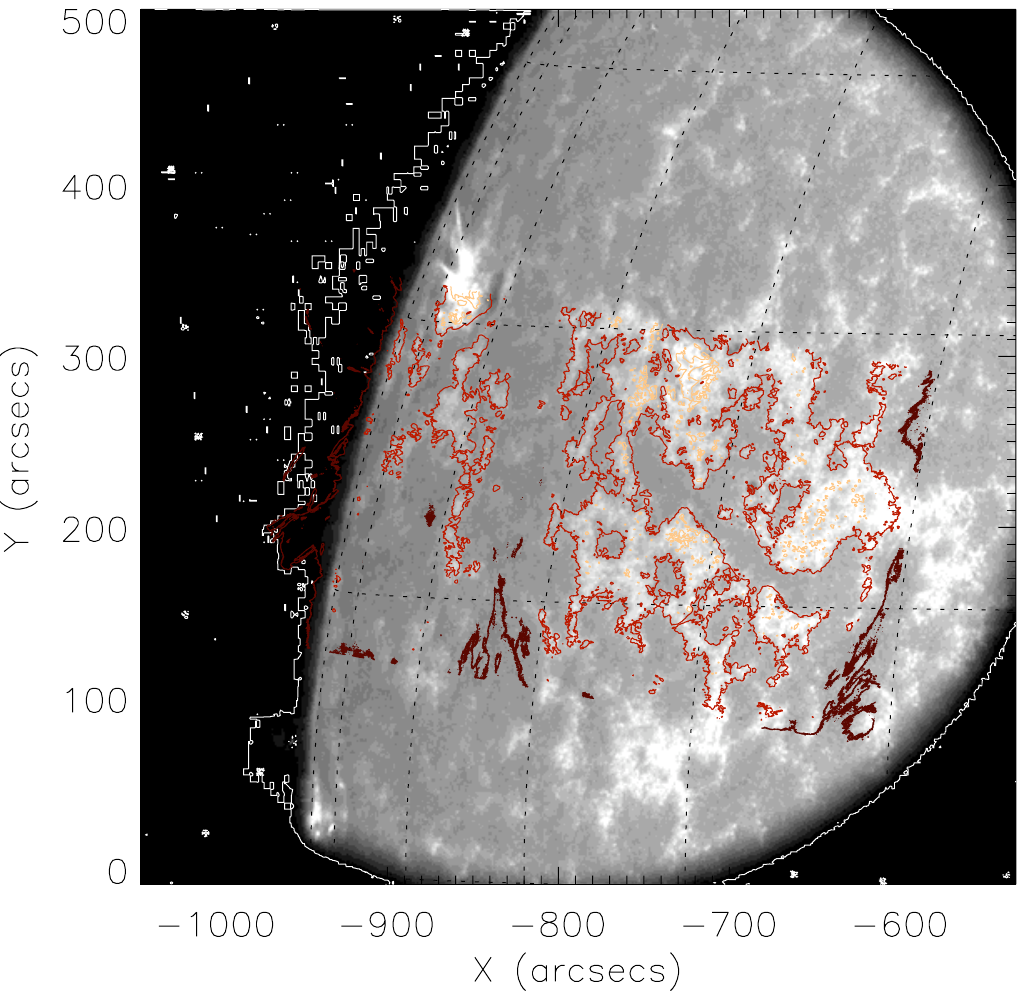} \includegraphics[width=0.53\textwidth]{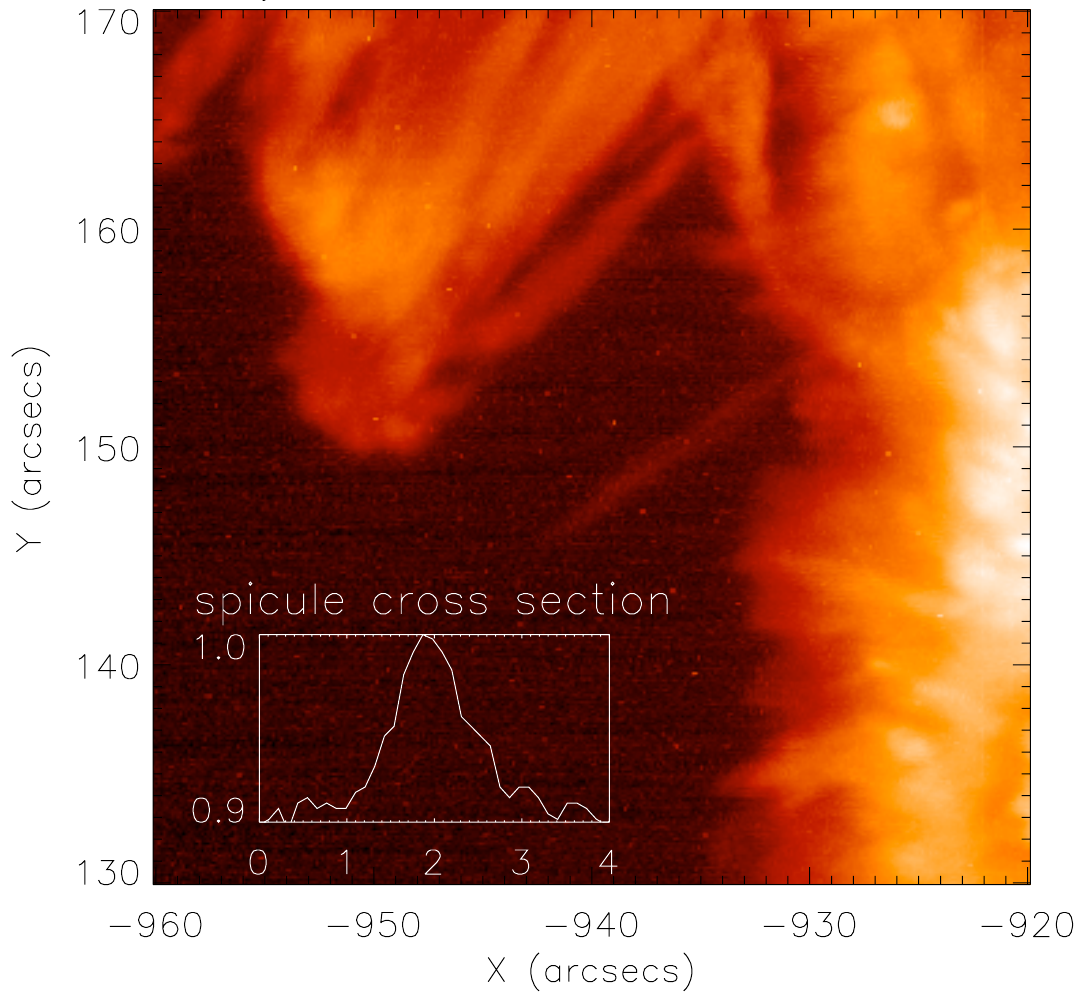}}  
\caption{\emph{Left: } Alignment of VAULT-TRACE Ly$\alpha$ images. We
   find the best correlation by optimizing the position, rotation and
   scale in both x-y directions. We derive a pixel size of
   $0.125\arcsec\times0.110\arcsec$. The TRACE Ly$\alpha$ was taken at
   18:18:54 UT (B\&W figure and white contour), the VAULT was taken at
   18:17:30 UT (three red contours). \emph{Right:} Estimation of the VAULT
   resolution using the thin spicule located at the center of the
   image. The median normalized cross section (in arcseconds) of the spicule
   along its length (plot inset in figure) is fitted with a gaussian,
   which leads to a FWHM $\approx0.49\arcsec$ as an upper
   limit. }\label{fig:align}\label{fig:resolution}
\end{figure}

\subsection{VAULT-II Observations \label{s:obs} }

VAULT-II obtained 21 images from 18:12:01 to 18:17:47~UT with a
cadence of 17 seconds. The integration time was 1 second for all
frames except for a 5-sec image (the 2nd in the series, not shown
here). The target was an old active region complex near the east limb
which included NOAA regions 9997-9999, Quiet Sun, filaments, plage and
the limb.

Figure \ref{fig:composite} is a composite image of all VAULT-II
frames.  The composite field of view (FOV) covers nearly 10\% of the
total visible solar disc area. To investigate possible center-to-limb
variation \cite{1956ApJ...124..580M} we have calculated the radial median
intensity of non-active region areas (excluding plage region,
prominences, flaring regions). We do not find any significant
center-to-limb gradient in agreement with \citeauthor{2008A&A...492L...9C} (2008).

The VAULT flight was supported by several other instruments. All
corresponding data (see table \ref{tab:JOP}) are available online or
per request.  

All VAULT data are publicly available online in FITS format and
compatible with \emph{SolarSoft} mapping routines. The images are
interaligned, the dark level is subtracted and an \emph{ad-hoc}
synthetic flat-field is also created and provided with the data, but
not applied on the online set. The flat-field is generated by
retrieving the median (in time) pixel value as the solar image moves during the
observations. It therefore accounts for flatfield and scattered
light. Intensities are left in DN.

To improve the visibility of faint, small scale structures, we have
applied a wavelet enhancement technique
\cite{2008ApJ...674.1201S}. This method decomposes the image into
frequency components (scales). The frequency decomposition is achieved
by means of the so-called a-trous algorithm.  With this method we can
then obtain an edge-enhanced version of the original image by
assigning different weights to the different scales upon
reconstruction. We note that the aforementioned decomposition does
not create orthogonal components, and therefore the reconstruction
does not conserve the flux. It is also possible to retain the
low-scale information by adding a model background image. Both
processed data, with and without the model are freely available in the
VAULT website. The level 0.9 VAULT-II data, together with the
IDL-\emph{SolarSoft } routines, composite full field image, flat-field and
wavelets processed data are available under:
\verb=http://wwwsolar.nrl.navy.mil/rockets/vault/= 

\begin{table}[h]
   \begin{tabular*}{\textwidth}{p{0.2\textwidth}p{0.2\textwidth}p{0.4\textwidth}} 
      Telescope   & Channel & co-temporal time serie \\
\hline
     SOHO &MDI/EIT/CDS 	& EIT-304 \AA\, partial FoV \\
     TRACE& 171, WL,1600& 171 \AA, partial FoV \\
     BBSO & H$\alpha$,Ca,B$_{LOS}$,WL & H$\alpha$, partial FoV \\
     Kitt-Peak & Mgram & Photospheric magnetogram\\
     \hline
   \end{tabular*}
   \caption{Joint observing campaign supporting the VAULT-II launch}
   \label{tab:JOP}
\end{table}

\subsection{Spatial resolution \label{s:spatial}}
The rocket pointing accuracy was $\sim 1$ arcmin with exceptional
pointing stability of 0.25\arcsec peak-to-peak over 10 sec. To obtain
the solar coordinates, rotation relative to solar North, and pixel
size for the VAULT-II images we used TRACE Ly$\alpha$ images taken
only minutes apart from the VAULT images. Figure \ref{fig:align} shows
the alignment results. The resulting VAULT pixel size is
0.125\arcsec$\times$0.110\arcsec\ which is in excellent agreement with the
optical design expectations \cite{2001SoPh..200...63K}. During the
flight, a small thermal expansion of the spectrograph structure
relative to the primary mirror resulted in an apparent
  pointing drift which was variable but less than $\approx$3
  pixels/second. If it was uniform during the flight, this drift would
  place a conservative lower limit on the instrument resolution of
  $\sim0.75\arcsec$ (0.375\arcsec/pixel). However, we could visually
  identify smaller structures (always in absorption on disk) in
  several images. To better estimate the actual image resolution we
measured the FWHM of the smallest structure we could find in the
images. We used the median cross section along the 110-pixel length of
the thinnest spicule, located at the center of Figure
\ref{fig:resolution} (right), and fitted it with a Gaussian. The
$FWHM\approx2.35 \cdot\sigma$, where $\sigma$ is the standard
deviation of the fitted gaussian profile. This leads to a VAULT-II
resolution $0.49\arcsec$. However, the upper limit may be dictated by
opacity effects rather than instrumental ones.

 The photometric calibration of the instrument was originally
 determined from the observations during its first flight in May
 1999. The calibration factor from digital units (DNs) to intensity
 (ergs\, s$^{-1}$\, cm$^2$\, sr$^{-1}$) was deduced by comparing the average
 emission (in DNs) of an area of the Quiet Sun to the Quiet Sun
 intensity obtained by \citeauthor{1974ApJ...187..369P} (1974). This was a reasonable assumption
 since both observations were made at similar phases of the cycle; the
 majority of the VAULT I field of view contained Quiet Sun and the
 \inlinecite{1974ApJ...187..369P} measurements are well calibrated ($\sim20\%$).

 A comparison to SUMER Ly-$\alpha$\ observations was made to improve
 on the radiometric accuracy of our measurements. The first issue was
 the spectral purity of the signal. The VAULT gratings transmit solar
 light in the range of $1140-1290$\AA. In this range there are only
 few relatively bright lines, the brightest of which is \ion[Si iii]
 at 1206.51\AA. SUMER Quiet Sun spectra show that almost all
 ($\sim95$\%) of the emission in this range comes from the Ly-$\alpha$
 (assuming a rectangular filter). Concerning the spectral purity on
 the full range above 120~nm, our calculations show that the signal
 should be 70\% pure.  
\begin{figure}
 \centerline{\includegraphics[height=0.5\textwidth]{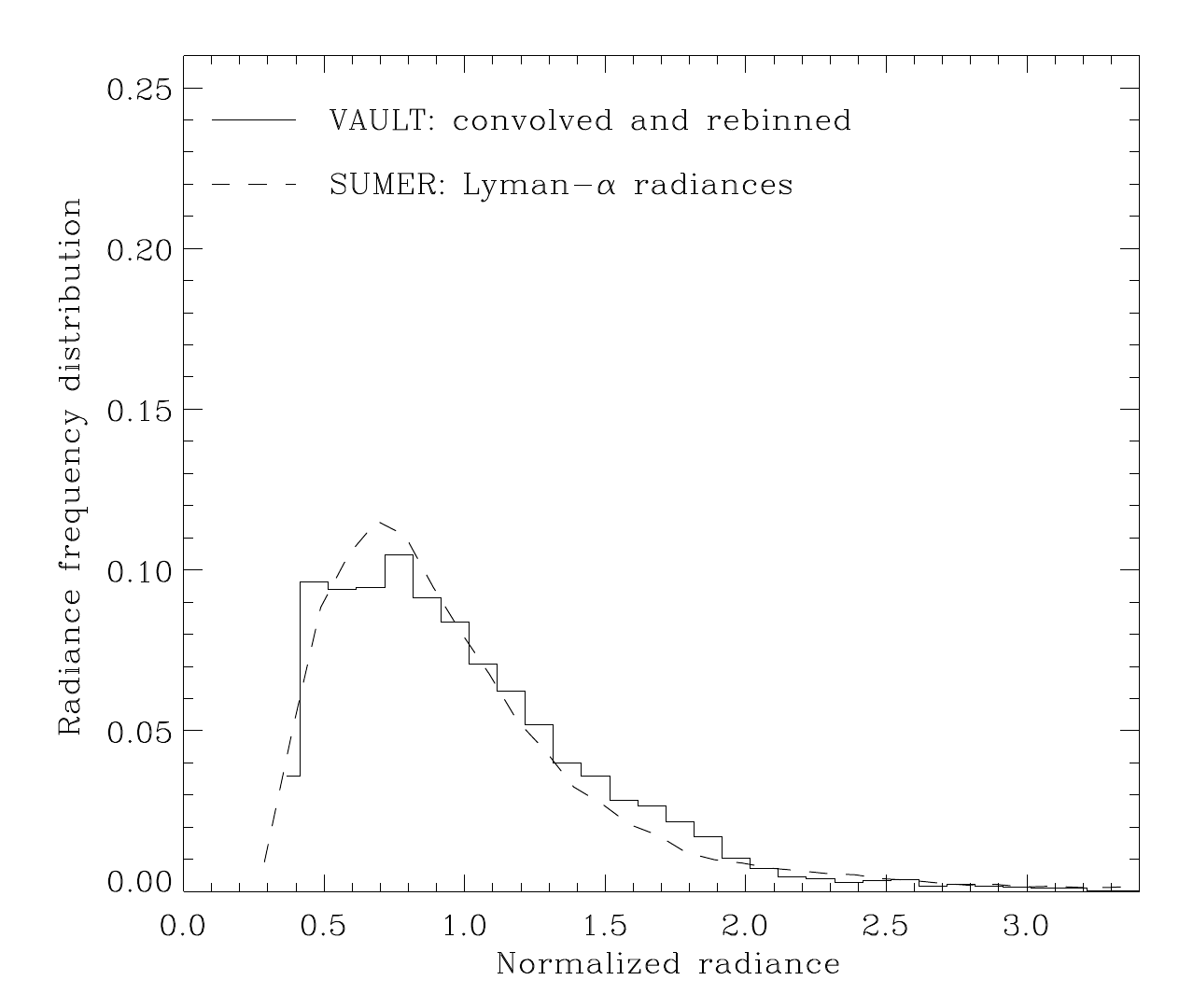}}
 \caption{Normalized \ion[H i] Ly$\alpha$ distributions as obtained
   from SUMER data (dashed line) and the VAULT data (solid line) rebinned and
   convolved to match the SUMER spatial resolution after subtracting
   a background signal level of 80 DN s$^{-1}$ (see section~\ref{s:spatial}). }\label{fig:sumer}
 \end{figure}

 Since the SUMER instrument has a photon counting detector with no
 dark signal, there is no background to be removed from the SUMER
 data. To establish the comparison with SUMER we assumed that the
 normalized radiance frequency distributions over quiet-Sun areas
 produced with data from the two instruments should be equal or very
 similar to each other. To account for the different spatial
 resolution we have also computed the radiance frequency distributions
 after convolving the VAULT data with a 2-D Gaussian function (of 12
 pixel=1.5$''$ FWHM, equal to the SUMER spatial resolution) and
 binning over $8\times8$ pixels to yield the SUMER pixel size of
 $\approx1''$.  The comparison revealed that a low level signal of
 about $\sim80$ DN s$^{-1}$ needs to be removed from the VAULT images
 to bring them in accordance to the SUMER measurements
 (Figure~\ref{fig:sumer}. After a careful examination of the VAULT-II
 images, we found a noise pattern of $83\pm20$ DN s$^{-1}$ which is
 variable from image to image and cannot therefore be removed with the
 dark current subtraction. We have traced the source of the noise to
 interference from a faulty ground when the payload is switched to
 battery power.

 The final step is a comparison of the Quiet Sun level in our
 images with an average Quiet Sun radiance measured at Earth as we did
 for the first flight. For the VAULT Quiet Sun level we used the peak
 of the histogram of the image intensities (in DN s$^{-1}$) minus the 83
 DNs of the background signal. The Quiet Sun level was $217\pm20$
 DN s$^{-1}$. The SUMER average Ly$\alpha$ radiance on the Quiet Sun in
 2008 was $73\pm16$ W m$^{-2}$ sr$^{-1}$. This is well within
 uncertainties with the \inlinecite{1974ApJ...187..369P} measurement
 of $78\pm16$ W m$^{-2}$ sr$^{-1}$. We adopt the latter value for
 consistency with our VAULT-I results and because it was obtained
 about two years after maximum and may better compare with our 2002
 data. In this case, we derive a calibration factor of
 1 DN s$^{-1}$ = $0.359\pm0.081$ W m$^2$ sr$^{-1}$.

 \begin{figure}
 \centerline{\includegraphics[height=\textwidth]{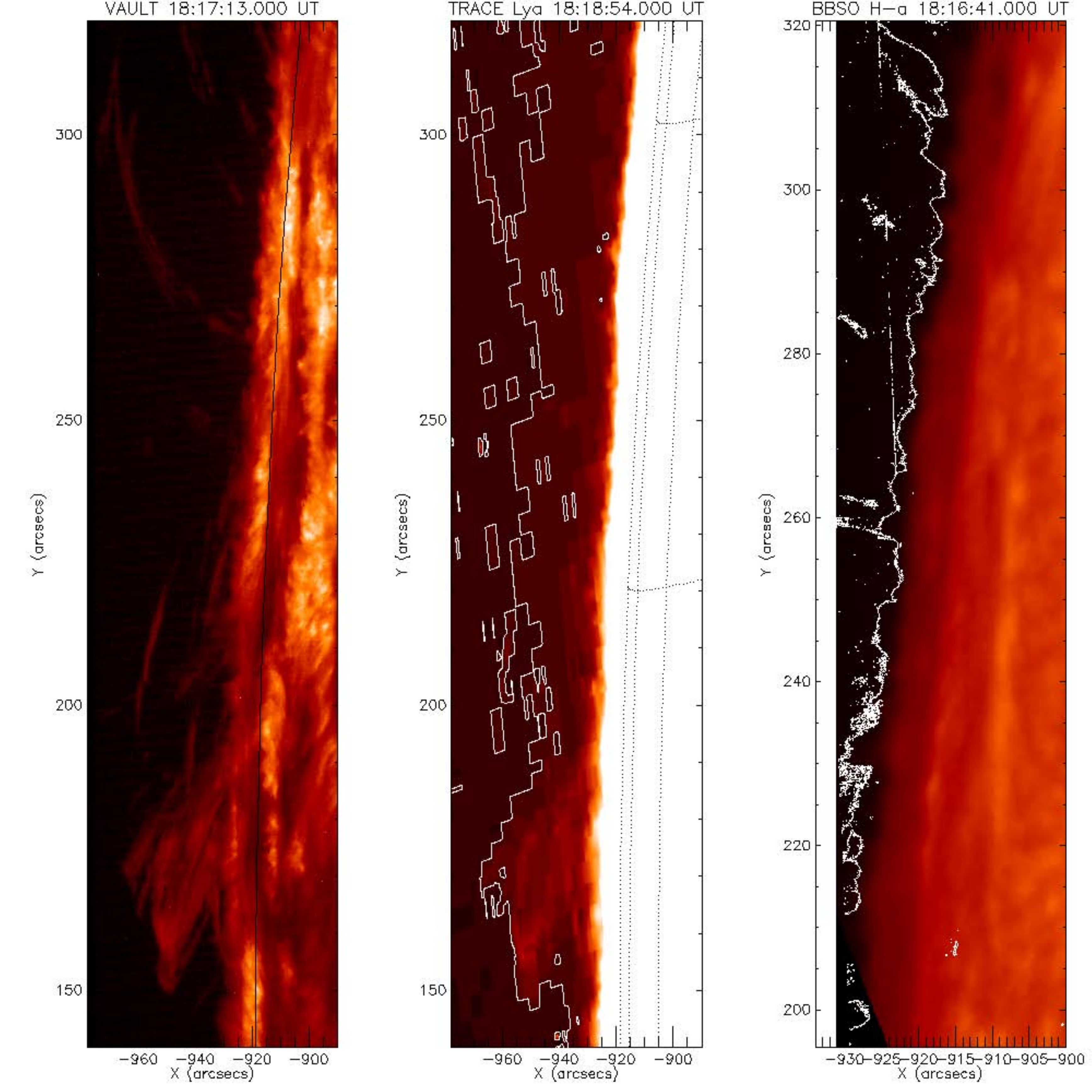}}
 \caption{VAULT-II, TRACE Ly$\alpha$, and BBSO H$\alpha$ comparison of
   the solar limb. All intensities are individually trimmed and scaled
   to emphasize the fainter details. \emph{Left:} VAULT-II after
   alignment and calibration. The black contour follows the visible
   limb, as aligned with the TRACE White light channel. \emph{Center:}
   TRACE Ly$\alpha$. The black grid denotes the photosphere. The
   prominence is much fainter than in the VAULT images but it is still
   visible. The white contour close to noise level helps to correlate
   with VAULT. \emph{Right:} H$\alpha$ channel from ground-based
   BBSO. The BBSO FOV is slightly smaller than the other
   instruments. The dotted contour denotes the VAULT edge. Note that
   the H$\alpha$ spicule heights are up to $\sim 2\arcsec$ shorter
   than in Ly$\alpha$.  }\label{fig:offlimb}
 \end{figure}

 \begin{figure}
 \centerline{\includegraphics[width=\textwidth]{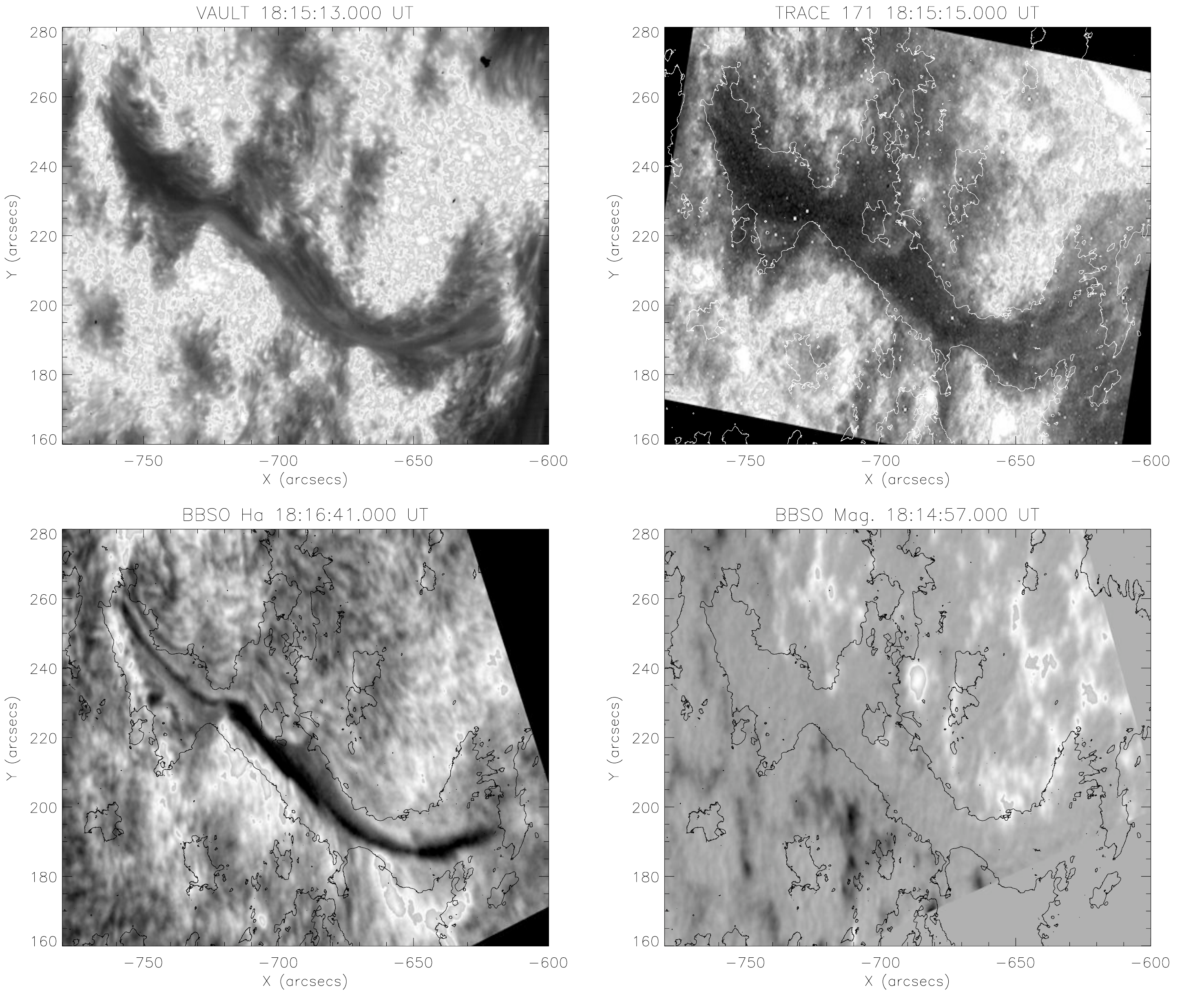}}
 \caption{Prominence as seen in almost simultaneous observations with
   various instruments. The contours mark the outer envelope of the
   Ly$\alpha$ prominence. The field of view is the same. \emph{Top
     left}: VAULT Ly$\alpha$.\emph{Top right}: TRACE
   171\AA. \emph{Bottom right}: SOHO/MDI photospheric
   magnetogram. \emph{Bottom left}: BBSO H$\alpha$ center.}\label{fig:4promin}
 \end{figure}

\section{The interpretation of the Ly$\alpha$ emission \label{s:ly}}

The hydrogen Lyman-$\alpha$ line, the strongest line of the solar
spectrum, is a $1s~^2S_{1/2}$ - $2p~^2P_{1/2,3/2}$ doublet resonant
line at 1215.67\AA. The FWHM of the line core is very broad ($\sim 1$\AA\
) due to Stark and Doppler broadening and the high optical thickness.
The line center probably forms in the lower TR ($\sim 40000$ K;
\opencite{1981ApJS...45..635V}) while the wings form in the
chromosphere ($\sim 6000$ K) by partial redistribution of the core
emission.  Thus, the Ly$\alpha$ line plays a critical role in the
radiation transport in the chromosphere/TR interface.  Below 8000 K,
model calculations show that the line is very close to detailed
balance.  For temperatures between approximately 8000 and 30000 K, the
dominant energy loss is through Ly$\alpha$ emission.  For temperatures
higher than about 30000 K, Ly$\alpha$ is transparent
\cite{2004A&A...413..733G}. The physics of this line have been
explored in a number of papers \cite{1981ApJS...45..635V,1986A&A...154..154G,1995ApJ...442..898W,2002ApJ...572..636F,2004A&A...413..733G} , and the average full-Sun line profile and its
variation over the solar cycle has been measured by the SUMER
instrument \cite{2004cosp...35..510L} but most deal with the spectral
characteristics and are of more interest to spectroscopic analysis. On
the contrary, VAULT data consist of the integrated line intensity over
a wide bandpass which includes contributions from other lines such as
\ion[Si iii], \ion[N i], \ion[N v], and \ion[C iii]. Because of the complexity of the line, model
calculations are the easiest way to interpret imaging
observations. Past analysis was based on plane parallel radiative
transfer models using the Ly$\alpha$ contrast (the ratio of the
Ly$\alpha$ emission of a structure relative to the average Quiet Sun)
to derive estimates of pressure and temperature within the observed
structures \cite{1982SoPh...75..139B,1986A&A...167..351T}. Recent
computational and theoretical improvements have enabled the
calculation of the emission from models with more realistic
cylindrical geometries \cite{2004A&A...413..733G} and therefore direct
comparison with observed Ly$\alpha$ intensities
\cite{2006ESASP.617E..63G,2007ApJ...664.1214P}. However, calculations from the latter
models remain time-consuming and difficult to apply over the wide
range of structures seen in the VAULT images. Since the scope of our
paper is to present a broad overview of the Ly$\alpha$ atmosphere, we
return to the plane parallel assumption and adopt the approach of
\inlinecite{1986A&A...167..351T} to estimate physical paramaters for
the structures in our images.  More careful analyses of specific
features will be undertaken in the future.

The calculations in \inlinecite{1986A&A...154..154G} require the
calculation of the ratio of the intensity of a given structure over
the average intensity over the solar disk or ``Ly$\alpha$ relative
intensity'' (LRI). Since we do not have full disk images in Ly$\alpha$
we cannot compute directly a solar disk average. However, the disk
emission is dominated by the Quiet Sun (Figure~5 in
\opencite{1974ApJ...187..369P}) and we therefore need only to
calculate the Quiet Sun level. Thanks to the large FOV, the VAULT
images contain large Quiet Sun areas. So, we use the median of the
lower part of the FoV $\sim(x\in[-550,-400],\forall y)$ in
Figure~\ref{fig:composite} as the "Quiet Sun" level. We then calculate
the LRI range for several representative features. 

The results are shown in Table~\ref{tbl:meas}.  The corresponding
pressure, temperature and optical thicknesses derived from
\inlinecite{1986A&A...154..154G} are also included. The numbers
suggest that most solar structures are optically thick in Ly$\alpha$
even at temperatures departing significantly from chromospheric ones
($\geq 10^4$ K). Quiet Sun emission seems to arise at the chromosphere
while plage, prominence and offlimb structures have lower TR
temperatures and are presumably located at larger heights. These
results are in agreement with the earlier measurements of
\inlinecite{1986A&A...167..351T} except of the minimum LRI
values. \inlinecite{1986A&A...167..351T} reported values as low as
0.05 but do not observe LRI below about 0.2 anywhere but at the edges
of offlimb loops. The difference is most likely due to higher
sensitivity and spectra purity of the VAULT instrument which should
increase the detected counts of the fainter structures and minimize
the continuum contribution to the Quiet Sun levels relative to past
instruments. The faintest structures (LRI $\sim 0.2$) seen in the
VAULT images are long, thin strands seen in absorption against the
network. These strands are also the smallest resolved structures with
the lowest temperatures (Table~\ref{tbl:meas}). They are very similar
to chromospheric filaments but they do not seem to be associated with
any large scale structure. Their origin is currently a mystery but they could be cooling loops. The best
candidates for optically thin emission are the offlimb loops seen in
the northeastern edge of the VAULT FOV. The observed LRI range of
0.4--0.5 could be consistent with either chromospheric ($<10^4$ K) or
TR emission. These loops were not detected in the BBSO H$\alpha$
images and thus we selected the higher temperature solutions (T$\sim
3-4\times 10^4$K) for them. 

\begin{table}[h]
\begin{tabular}{llllll}
Structure & Intensity & Radiance &
Opt. Depth & T & Pressure \\ 
 & [LRI]$^\dagger$ & $[10^{12} {ergs\over cm^2 s\ sr}]$ &$Log$ & [$10^3$K] & [dyn/cm$^2$]\\ 
\hline   
Quiet Sun        & 0.5 --- 5  & 3.3 --- 32.5 & 4 --- 5& 8 --- 10   & 0.1 --- 1 \\ 
Quiet Sun Prom.    & 0.2 --- 1.4& 1.8 --- 9.5& 6 --- 3& 7 --- 9 (20)*& 1 (0.1)* \\
Plage            & 5.7 --- 12 & 37.5---75.0  & 4    & 10---13  & 1       \\
Plage Prom. & 1---5   & 6.7---32.5& 3---0& 8---40   & 0.1---1 \\
Offlimb Prom.    &0.8---1.1 & 5.8---7.8& 3---0& 15---80  & 0.1---1  \\
Offlimb Loops    &0.4---0.5 & 2.8---3.8& 0   & 30---40  & 0.1       \\
\end{tabular} 

\caption{Qualitative plasma diagnostics for several types of
  structures. See Section \ref{s:ly}.
\newline* Likely to have reduced optical thickness. High values
reflect underlying plage. 
\newline $\dagger$ Ly$\alpha$ relative intensity (LRI). LRI=1
represents median of Quiet Sun region.} 
\label{tbl:meas}
\end{table}
Table~\ref{tbl:meas} serves as a concise description of the physical
parameters of Ly$\alpha$ structures and we will refer to it in our
subsequent discussions of individual features starting the contribution 
of each of these features to the overall Ly$\alpha$ intensity. 

\section{Sources of the Ly$\alpha$ Intensity \label{s:inten}}

 \begin{figure}
 \centerline{\includegraphics[width=\textwidth]{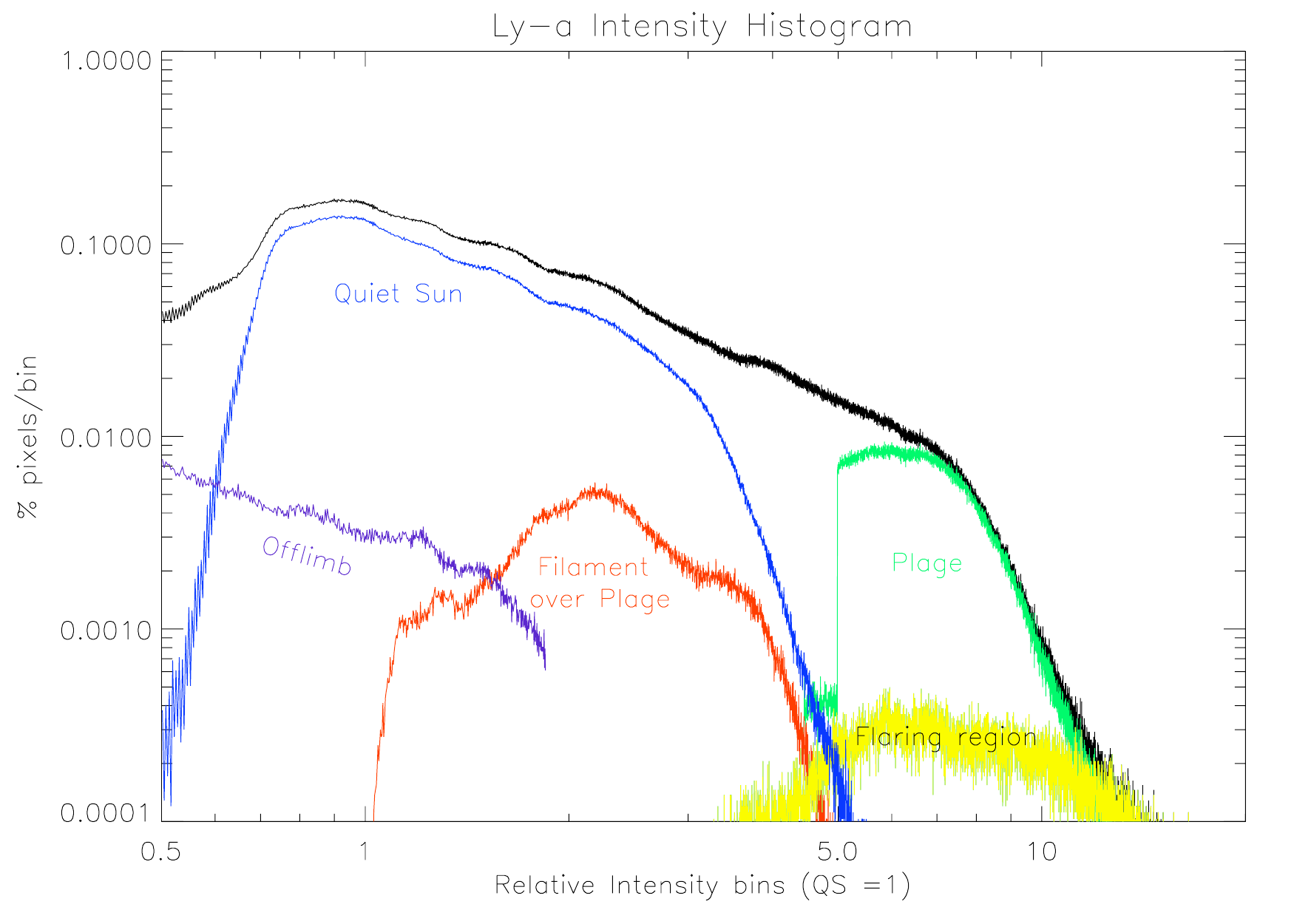}}
 \caption{Ly$\alpha$ emission histogram (black line).
   Different colors represent partial histograms from the labeled
   subregions. In particular we note that the "Quiet Sun" emissions
   spans one order of magnitude. Intensities are scaled to the median
   Quiet Sun level. Plot ordinates scaled to total number of data
   pixels. Cover area for each type (integral over the histogram
   curve) is: Total (black line): 100\%, Quiet Sun: 61\%, Plage:13\%,
   Filament: 2\%, Flaring region: 1\%, Offlimb: 1\%, Rest: 23\%. See
   text on Section \ref{s:dis} for more details.}\label{histogram}
 \end{figure}

 Ly$\alpha$ is a very optically thick line and results in both
 emission and absorption depending on the properties of the
 surrounding plasma. This interplay is at the region where the plasma
 starts to be dominated by the magnetic fields, creating a wide range
 of intensities. On the other hand, the strength and variability of
 the Ly$\alpha$ irrandiance has important effects on Earth because it
 affects the chemistry of the mesosphere (e.g., ozon layer) as well as
 the climate on longer time scales. Only the central part of the broad
 spectral profile of the solar Ly$\alpha$ emission is effective for the
 geo-environment. But there is a clear relationship between the
 central radiance of the solar Lya line and the total irradiance of
 the line \cite{2005Icar..178..429E}.To understand
 changes in Ly$\alpha$ irradiance we first need to identify the
 contributions of the various solar sources of this emission to the
 total Ly$\alpha$ irradiance.

 We attempt a first cut at this problem using our spatially resolved,
 calibrated images. As we discussed before, we are able to
 differentiate among Quiet Sun, Plage, Prominence over
 Plage, Offlimb and Flaring regions.  Figure \ref{histogram} shows the
 corresponding intensity histograms for each domains (color coded),
 relative to the overall histogram (black line). The values are
 constructed from the pixels inside each region, and considering the
 median value for each pixel in time (from Figure~\ref{fig:composite}):

 \emph{Quiet Sun (blue line):} We select a region around the lower
 right corner in Figure~\ref{fig:composite} as typical Quiet Sun. Based
 on this selection, the Quiet Sun covers 61\% of the pixels. We use
 the median value of the Quiet Sun as a normalizing factor. Normalized values
 inside this region, however, span from 0.5 to 5. The Quiet Sun exhibits a
 wide range of intensities, as it can be expected by the high optical
 thickness and strong structuring of the plasma. The low end of the
 histogram reaches the edge detection of offlimb prominences, while
 the high end reaches the plage levels. Scattered around this Quiet Sun we
 find several cases of localized brightenings which may be related to
 explosive events, which we discuss in Section~\ref{s:qs}.

 \emph{Plage (green line):} The central part of the VAULT FOV shows a
 bright plage. Following a similar method as for the Quiet Sun we find that
 the plage covers 13\% of the pixels, without considering the central
 overlying filament. Typical normalized intensities range from 5 to
 15. The only other contribution at these levels comes from the
 flaring region at the north edge of the image. This means that one
 approximation to the total solar Ly$\alpha$ irradiance can be
 obtained using the Quiet Sun level adding a multiplying factor $\sim$7 for
 the percentage of the disc corresponding to plages (which could be
 obtained from other lines like Ca).

 \emph{Filaments over plage (red line):} Our results show that the plasmas
 in the filaments over the plage are sufficiently opaque to
 reduce the observed intensity to Quiet Sun values. This
 particular filament blocks the central 22\% of the plage area.

 \emph{Offlimb (purple line):} The VAULT images contain several
 examples of limb structures, including spicules. As discussed later,
 we find higher heights for the spicules compared to H$\alpha$. Large
 overlying loops reaching projected heights of 60\arcsec\ can also be
 observed. The emission from these structures indeed shows Quiet Sun levels,
 down to our detection threshold for the histogram (0.5). It is likely
 that these structures are nearly optically thin, implying temperature
 $\gtrsim$30,000K. The large heights imply a dynamic state for these
 loops and they are probably associated with catastrophic cooling
 episodes studied previously with \textsl{TRACE\/}
 \cite{2001SoPh..198..325S}.

\section{Prominence and Filament Observations \label{s:prom}}

\begin{figure}
 \centerline{\includegraphics[width=0.8\textwidth]{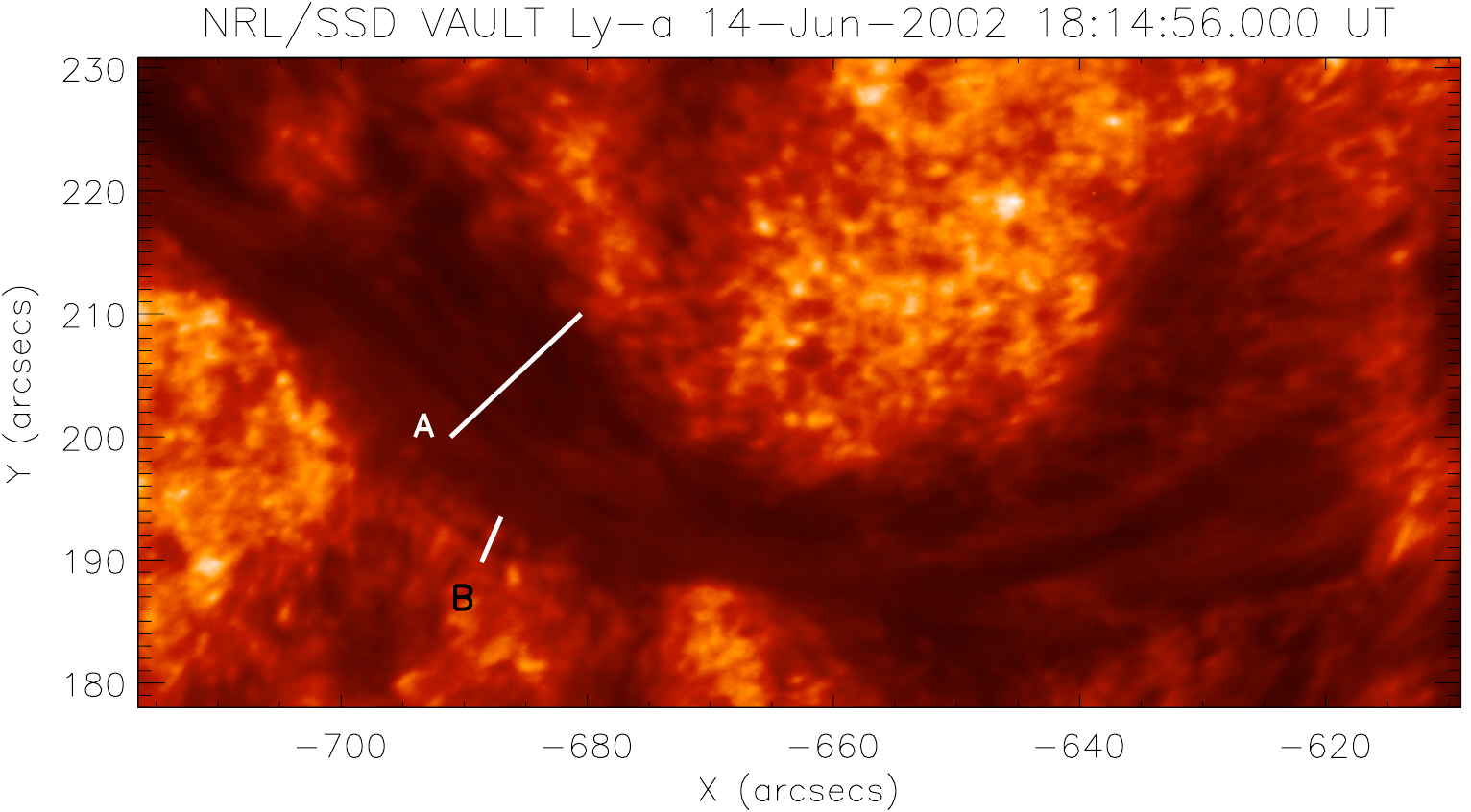}} \centerline{\includegraphics[width=0.9\textwidth]{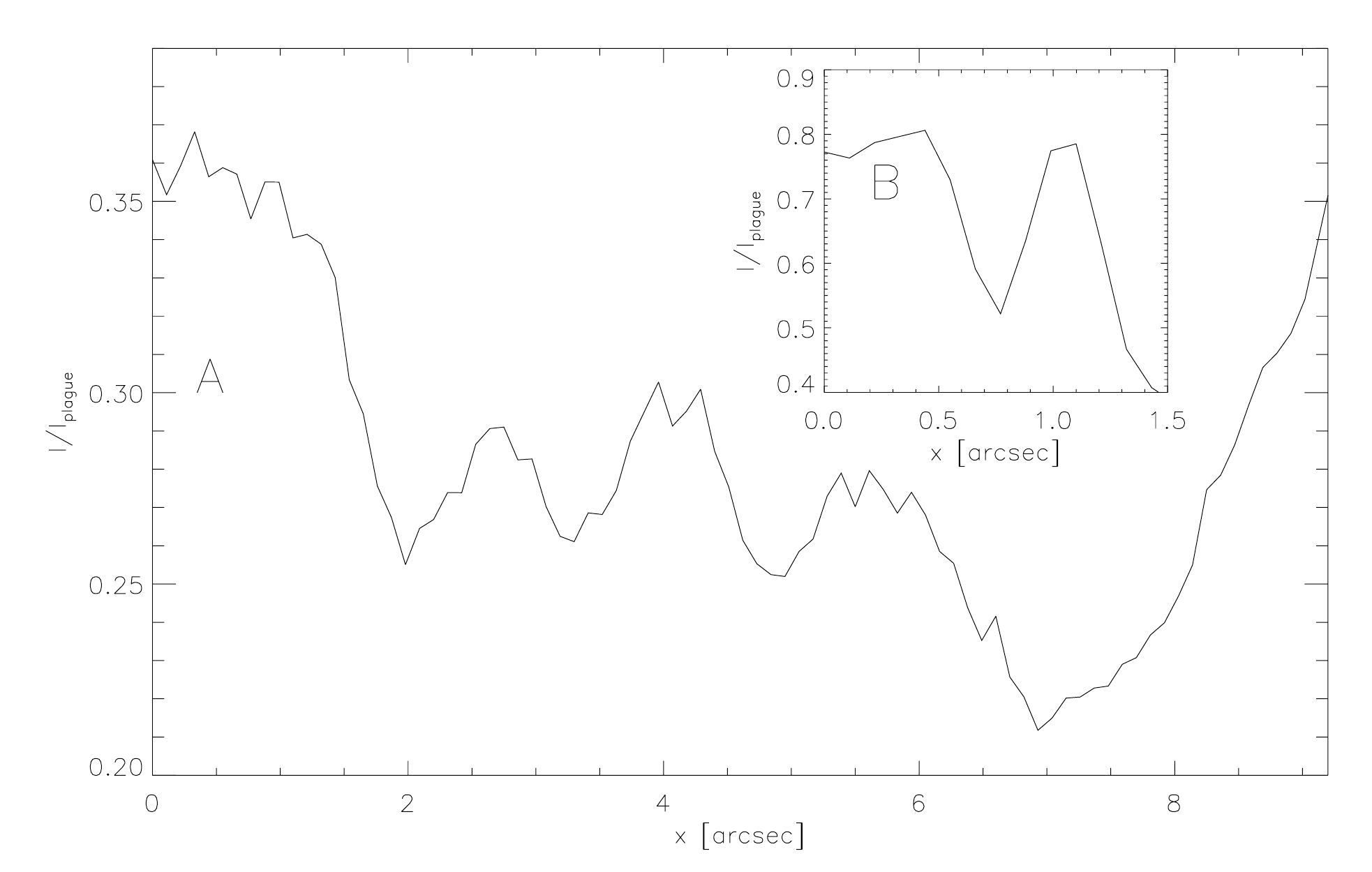}}
 \caption{Prominence shows reduced absorption threads along its axis
   below 0.5\arcsec. This supports a non-axisymetric
   prominence. \emph{Top}: Single VAULT frame, with labeled segments
   for the perpendicular cuts (labeled as 'A' and 'B'). Intensities
   are not scaled. \emph{Bottom}: Intensity profile perpendicular to the
   prominence axis for 'A' segment and 'B' detached thread (plot
   inset). Intensities are scaled to surrounding median plage
   value.}\label{fig:threads}
\end{figure}

The images contain a large number of filaments, filamentary structures
and a prominence and it is the first time that the fine scale
structure of the filaments is resolved in this wavelength. Figure
\ref{fig:threads} reveals a highly organized filament comprised of
parallel threads with little, if any, twist. No obvious twist is evident in
any of the other filaments as well.  The threads have a typical width of
around 0.5\arcsec or less, and are seen as intensity enhancement
profiles of about 5\%. Figure \ref{fig:threads} also shows a stable
and detached thread with a width reaching the instrument resolution
and 30\% absorption over the underlying plage. The filament is
further analyzed in \inlinecite{Millard:2009jk} where the comparison
with the H$\alpha$ observations suggests that Ly$\alpha$ traces the
cool outer plasma while H$\alpha$ originates from the coolest part of
the filament. There is also evidence for uneven absorption across the
filament axis. The northern side shows evidence of Ly$\alpha$
absorption while the southern side shows absorption only in the coronal
lines (171\AA\ ) consistent with the presence of a void or cavity
around the filament. The northern absorption could be understood as a
line of sight effect from low-lying absorbing plasma at the filament
flanks.   

The last panel in Figure~\ref{fig:offlimb} shows the size
discrepancy between Ly$\alpha$ and H$\alpha$ observations. Only a
small knot, $\sim5\arcsec $ width, of H$\alpha$ emission is visible
whereas the Ly$\alpha$ prominence extends for almost 50\arcsec.  

\section{Quiet Sun observations \label{s:qs}}
\begin{figure}
  \centerline{\includegraphics[width=\textwidth]{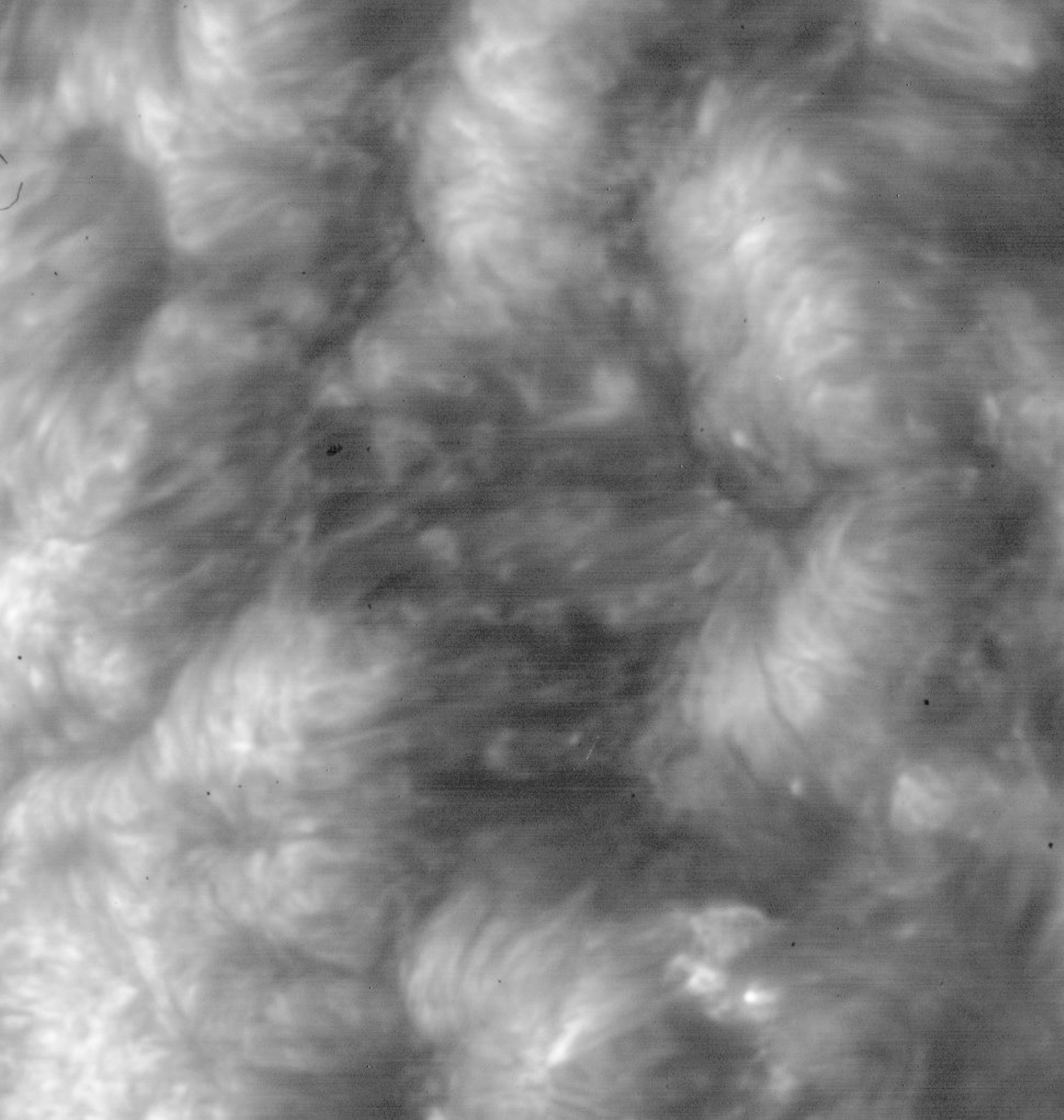}} 
  \caption{Detail of a supergranular cell in the Quiet Sun in
    Ly$\alpha$. The horizontal extent is 121\arcsec\ and the vertical
    is 117\arcsec\ and the field of view is centered at around (-500,
    250) in Figure~\ref{fig:composite}. These are the first spatially
    resolved images of the Ly$\alpha$ emission of a cell
    interior.} \label{fig:qs}
\end{figure}

The Quiet Sun has been the testing ground for the various theories and
concepts of the structure of the solar atmosphere. It is not
surprising then, that it is also the area where VAULT observations
have generated the most interesting results
\cite{2007ApJ...664.1214P,2008ApJ...683L..87J}. Earlier observations
showed that Ly$\alpha$ emission is concentrated along the
supergranular lanes in clumps with small loop-like extentions towards
the cell interiors. Faint emission without spatial structures was
detected at the cell centers. VAULT images, especially VAULT-I which
covered a much larger Quiet Sun area, resolved the spatial stucture in
the clumps along the supergranular boundaries (Figure~\ref{fig:qs}).
The Quiet Ly$\alpha$ Sun area shows groupings of filamentary plasma,
similar to the H$\alpha$ rosettes, with a typical diameter of $\sim
23\arcsec $. These rosettes show filamentary structure up to resolution
limit of the instrument, of about $0.4\arcsec$. This grouping in
rosettes is stable through the observations ($\sim6$ min) but shows
the presence of localized brightening events with a timescale variation
$60-120$ sec and sizes of a couple of arcseconds. The network
structures rise above the chromosphere about 7100~km or 10\arcsec\ as
seen in Figure~\ref{fig:offlimb}.  This measured value is consistent
with previously measured values of the height of the transition region
above the limb.  Their location at the supergranular cell boundary
uniquely identifies these loops as being the byproduct of convective
motion driving together magnetic fields at the edges of the
supergranular cell.

The outer areas consist of short loop-like structures while the
centers of the clumps have a more point-like nature. This morphology
is consistent with loops of progressively higher inclination towards
the center of the boundary. The obvious question is whether the more
extended Ly$\alpha$ loops are full loops or just the lower part of a
larger structure, possibly extending to higher
temperatures. \inlinecite{2007ApJ...664.1214P} applied an analysis
method used for coronal loops to a detailed Ly$\alpha$ emission model
and found that the short loops at the edges of the boundary channel
were consistent with full Ly$\alpha$ loops and therefore could account
for the ``cool'' loops predicted by models of the transition region
\cite{1986SoPh..105...35D}. However, the magnetic footpoints of these
loops could not be identified in photospheric magnetograms due to the
lower spatial resolution and reduced sensitivity of the MDI
data. Although these problems should not affect the larger loops,
their footpoints remain ambiguous. To address these problems,
\inlinecite{2008ApJ...687.1388J} decided to investigate the magnetic
origin of the extended Ly$\alpha$ loops using magnetic field
extrapolations. They found that the longer Ly$\alpha$ loops originate
near the boundary center and are more likely the lower extensions of
large scale loops that connect areas much more distant than the neighboring
cells. The extrapolations showed that the smaller loops at the edge of
the network lanes are indeed small scale loops supporting the
interpretations of \inlinecite{2007ApJ...664.1214P}.

\subsection{Cell Interior}
Another new observation from VAULT is the imaging of Ly$\alpha$
emission from the cell interiors for the first time. As can be seen in
the example of Figure~\ref{fig:qs}, the emission extends over the full
interior area and is structured in various spatial scales. The
emission is filamentary, optically thick with some apparent
dependence on the local radiation field. The associated time series
(movies available in the online VAULT archive) reveal significant
evolution in these structures, like flows and jets. The material
within the filamentary structures shows an overall motion towards the
network boundary similar to the motions of emerging magnetic field
elements in photospheric magnetograms and white light images. As
magnetic field of opposing direction accumulates in the boundary, it
is expected that some cancellation is taking place. Indeed, there are
a few cases where Ly$\alpha$ material appears to jet out from smaller
emission clumps creating a bright point. These events are never seen
in the cell center and could originate from magnetic reconnection
closer to the photosphere. Some examples can be seen along the column
at $-500\arcsec$ in Figure~\ref{fig:composite}. The limited resolution
of available magnetograms has not allowed us to locate the origins of
these jets.
\begin{figure}
  \centerline{\includegraphics[clip=true,trim= 4.3cm 0.7cm 5.2cm
    0.7cm,,width=0.25\textwidth]{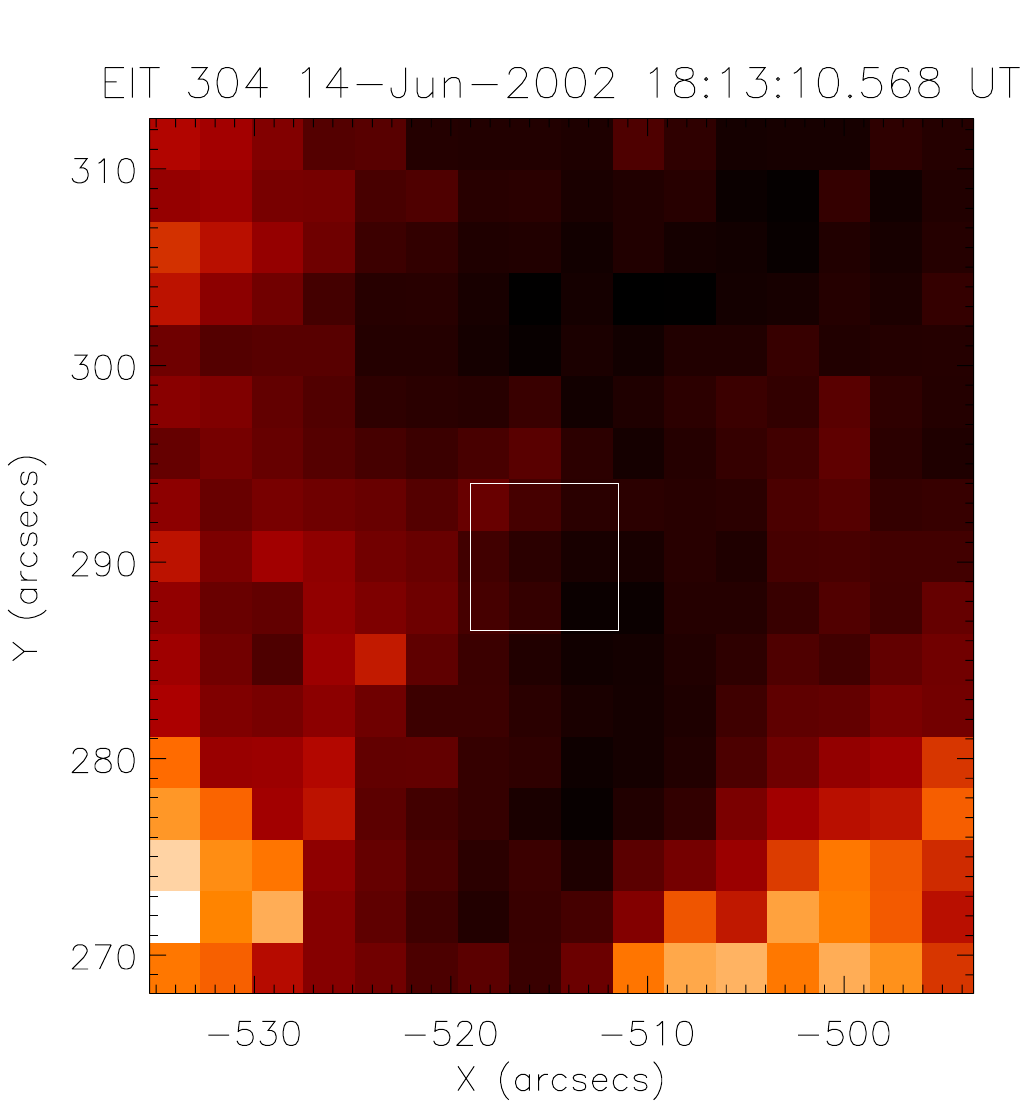}\includegraphics[clip=true,trim=
    4.3cm 0.7cm 5.2cm
    0.7cm,,width=0.25\textwidth]{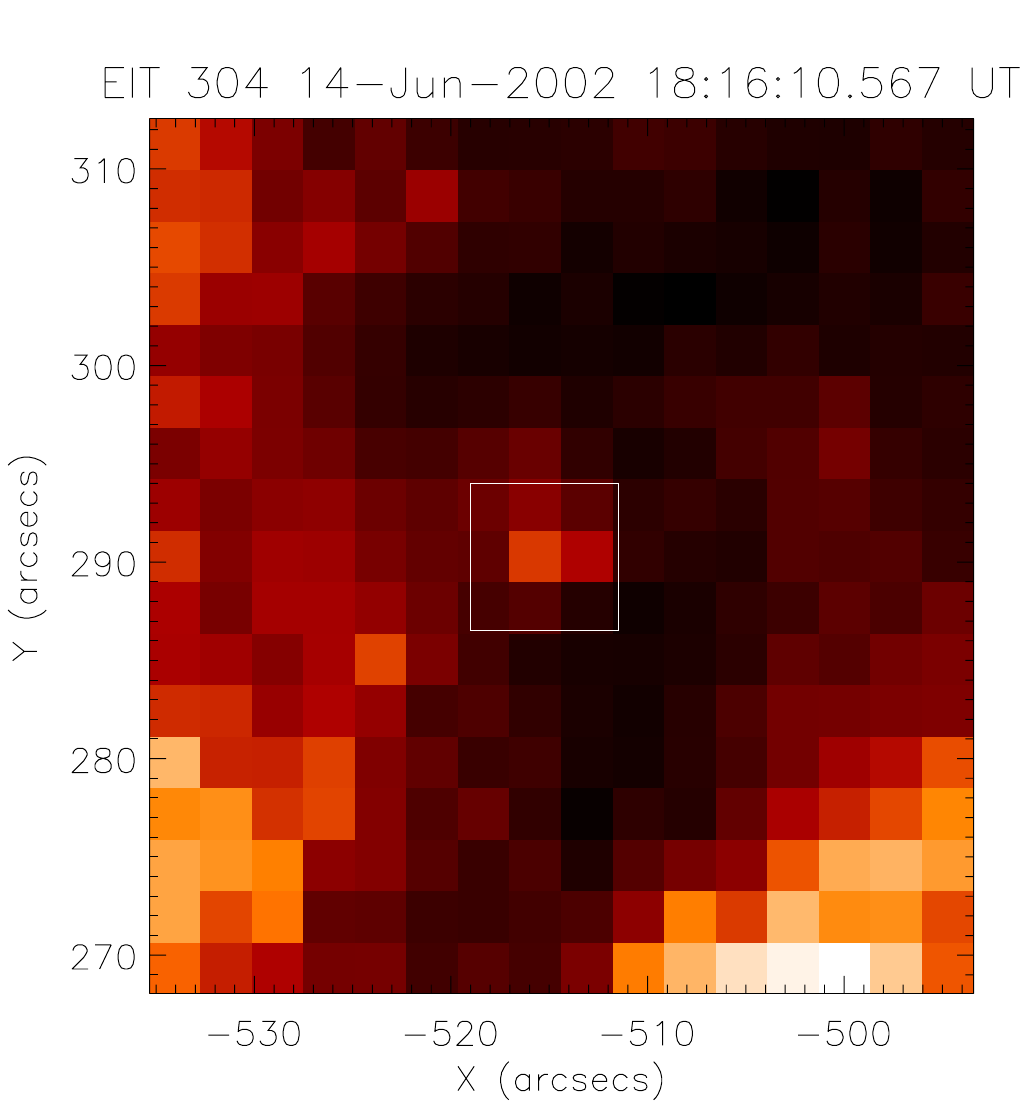}\includegraphics[clip=true,trim=
    4.3cm 0.7cm 5.2cm
    0.7cm,,width=0.25\textwidth]{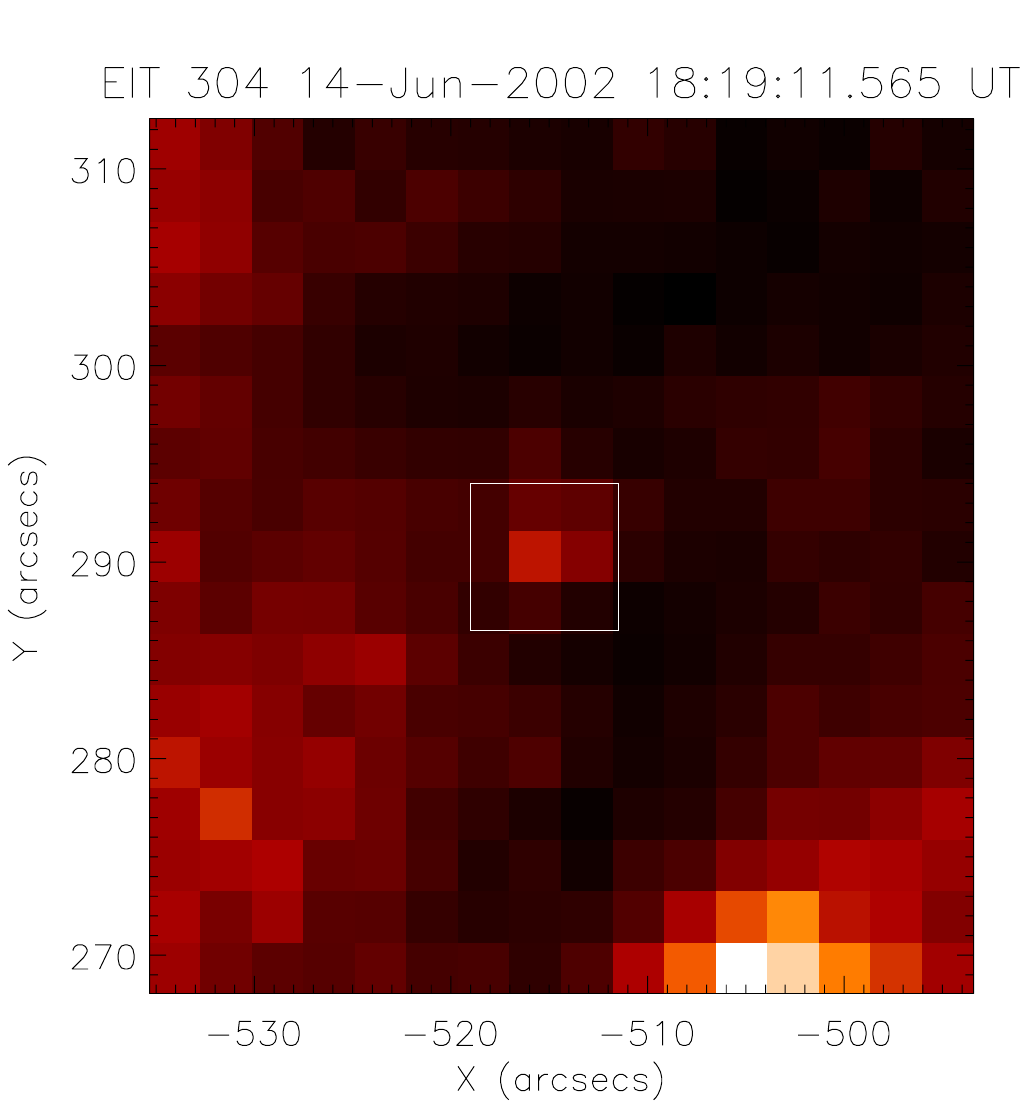}\includegraphics[clip=true,trim=
    4.3cm 0.7cm 5.2cm 0.7cm,,width=0.25\textwidth]{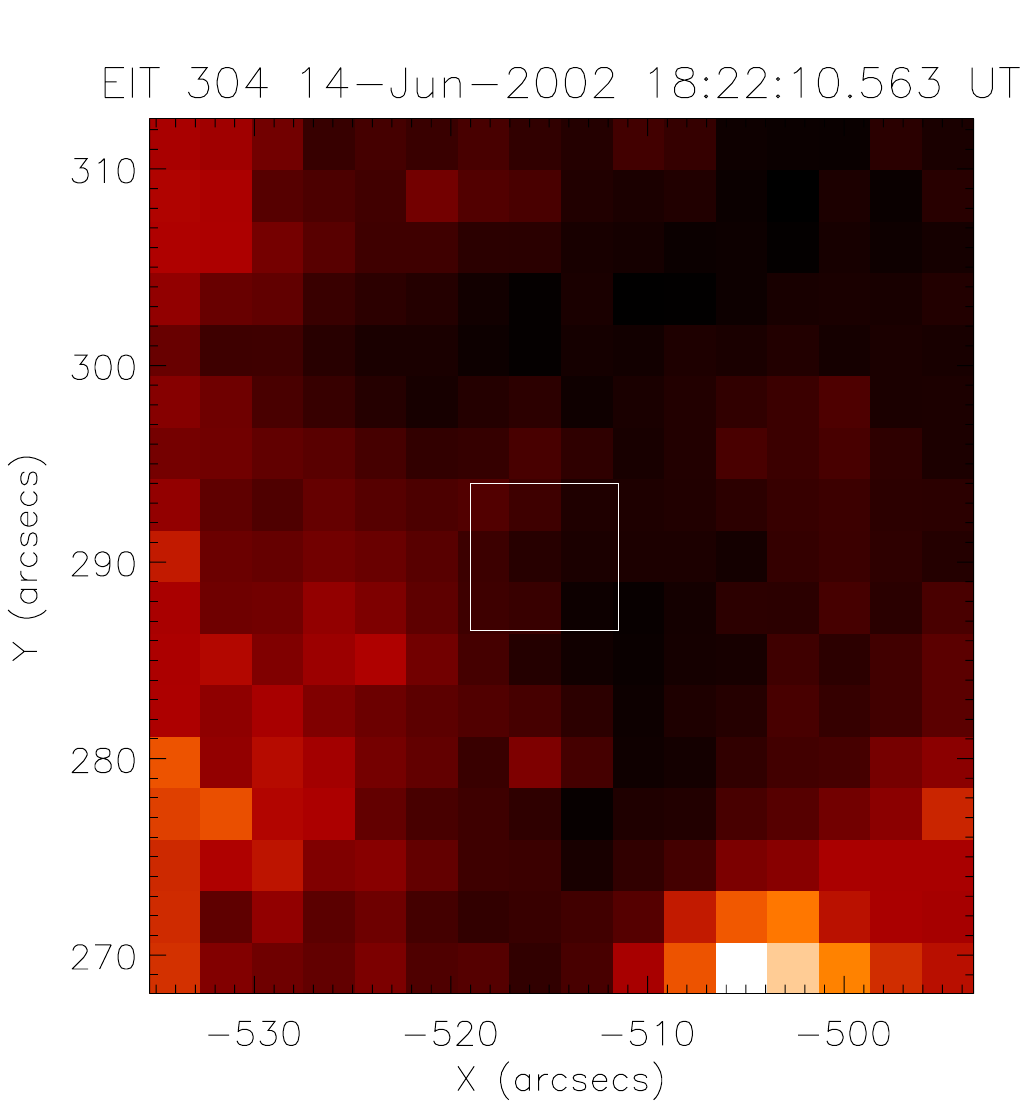}}
  \centerline{\includegraphics[width=\textwidth]{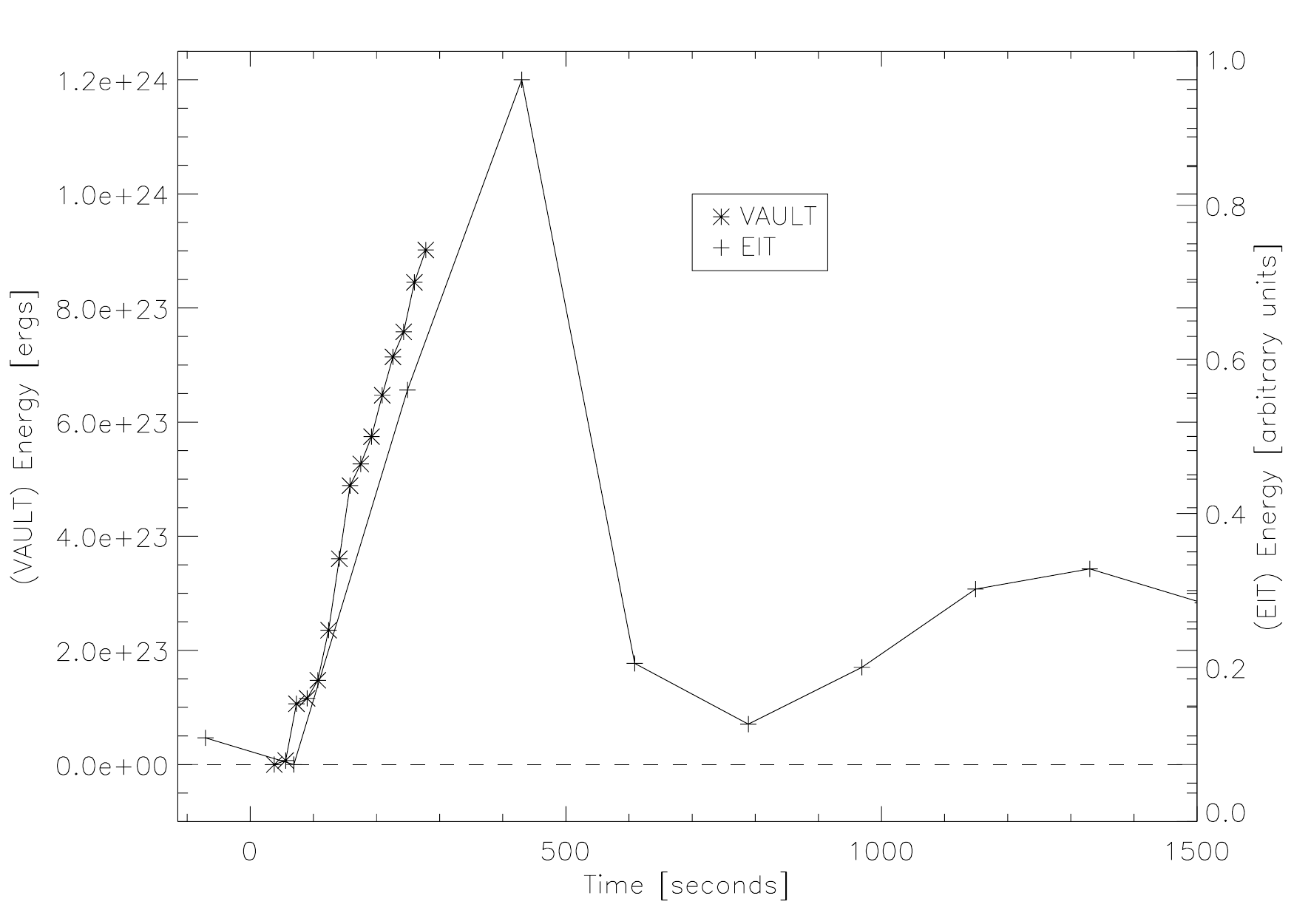}}
  \centerline{\includegraphics[width=0.5\textwidth]{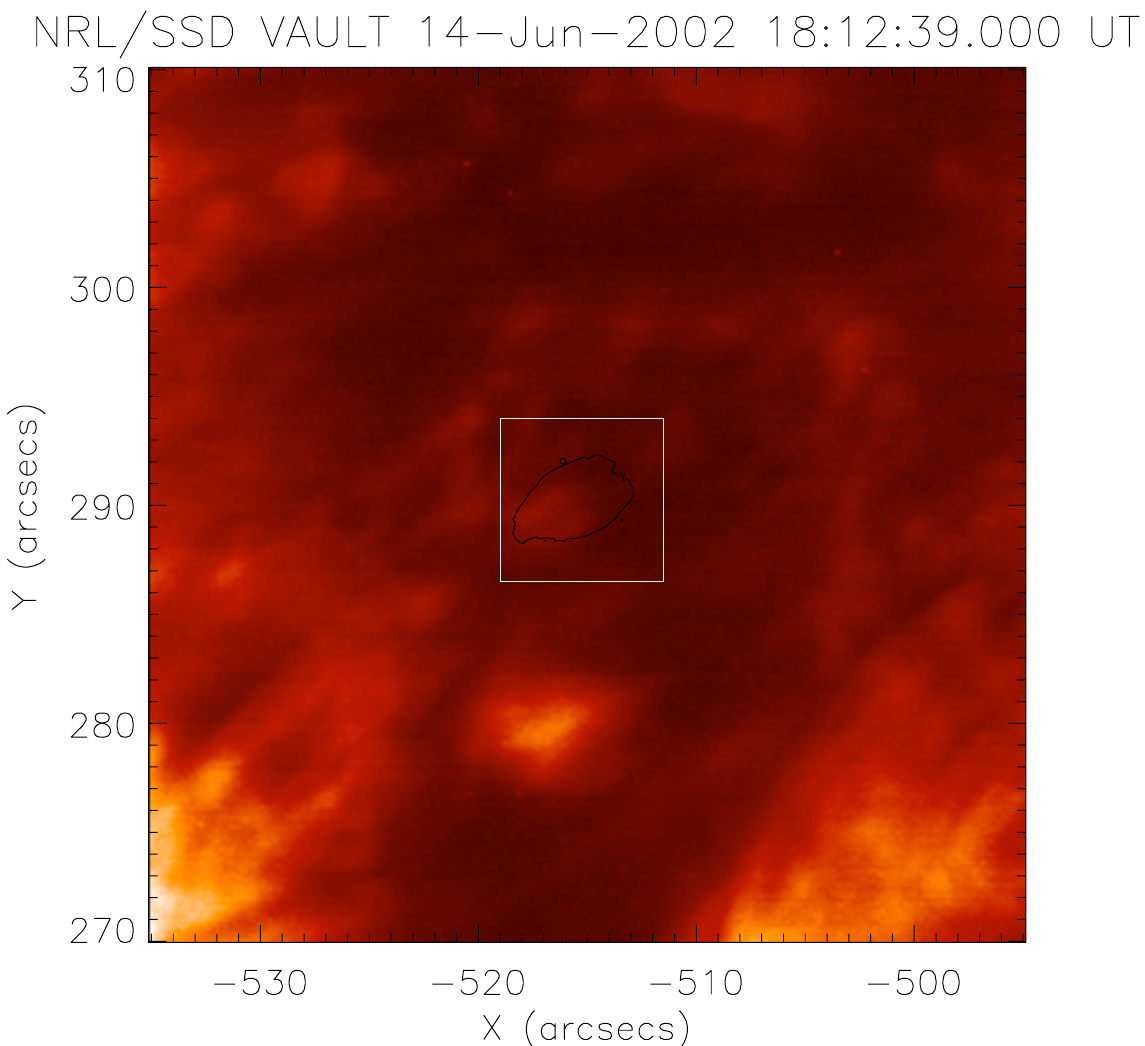}\includegraphics[width=0.5\textwidth]{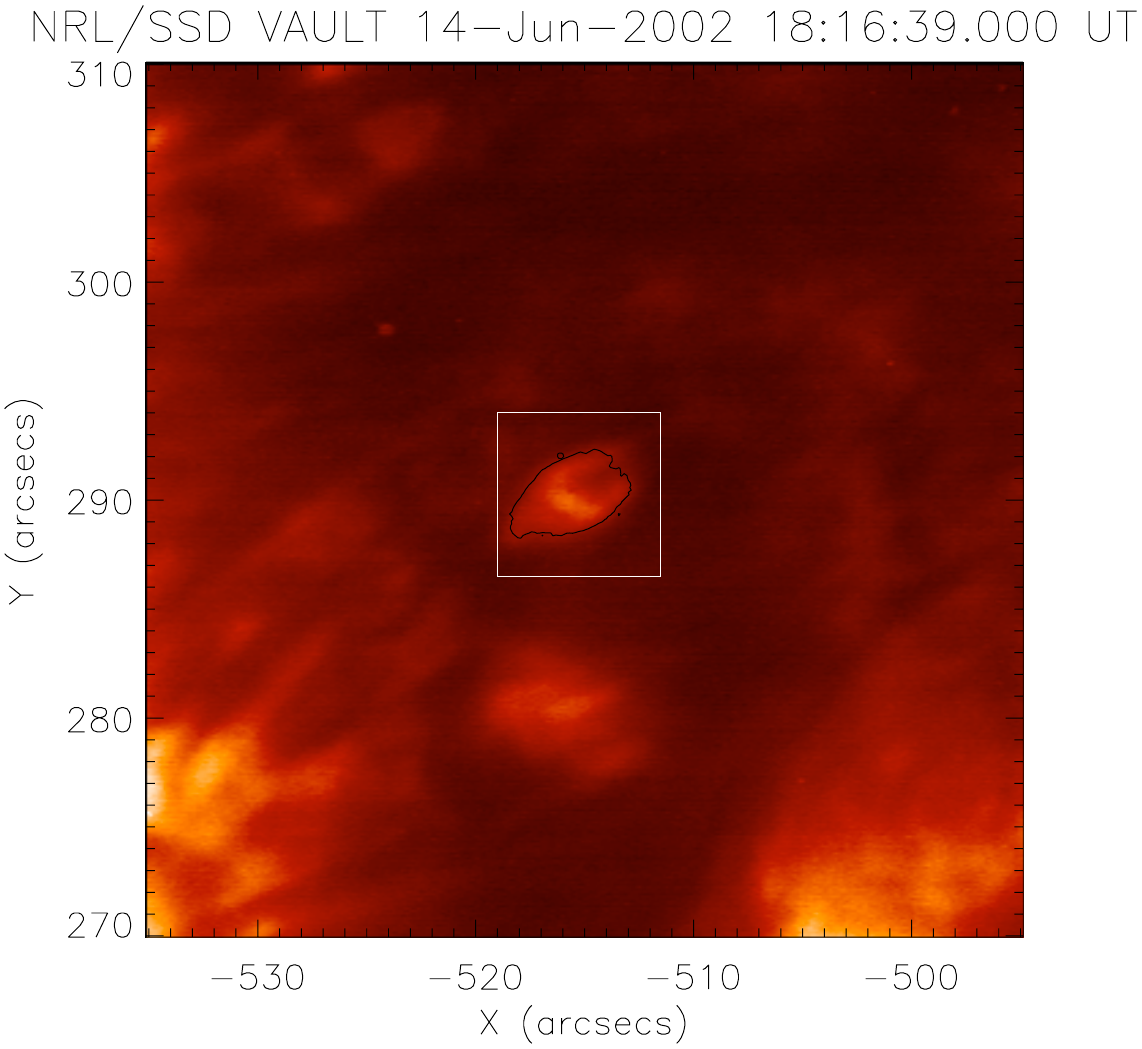}} 
  \caption{Microflaring event in the Quiet Sun detected in Ly$\alpha$
    and \ion[He ii]. \textsl{Top panels:\/} EIT \ion[He ii] images of the
    event. \textsl{Middle panel:\/} Comparison of the energy curves from VAULT to the EIT light curve. 
    The total counts within the black outline (VAULT) and within the white boxes (EIT) were
    used to calculate the curves. \textsl{Bottom panels:\/}
    Ly$\alpha$ images of the event at its initiation (left) and peak
    (right).  }\label{fig:blob}
\end{figure}
 
\subsection{Mircoflaring in the Quiet Sun}
Although the VAULT time series show continuous motions and brightness
evolution thoughout the full field of view, there are very few strong
enhancements that could qualify as flaring emission. The short
duration of the flight may be a reason for this but we were able to
isolate only $2-3$ events. Figure~\ref{fig:blob} shows an example
from a Quiet Sun feature which gives rise to a plasma jet rising
from the cell center. The brightening was detected by SOHO/EIT which
classifies it as a regular bright point. The event lasts for $\sim500$ s.

Since we have calibrated images, we could estimate the thermal energy of the
Ly$\alpha$ flaring under some assumptions.  We adopted equation (5) in
\inlinecite{2002ApJ...568..413B}  
\begin{equation}
E_{th} = 3k_BT\sqrt{EM V}
\end{equation} 
where the energy $E_{th}$ corresponds to Ly$\alpha$ plasma at
temperature $T$ and emission measure, $EM$ integrated over volume
$V$. We assumed $T=2\times10^4K$, $V=\dA\dl$, $dA=21\arcsec\times 21\arcsec$ area, and
$dl=0.5\arcsec$ equal to the mean free path of a Ly$\alpha$ photon for
optically thick emission. For the estimation of $EM$ we adopted the
calculations in \inlinecite{2001ApJ...563..374V} but used the updated
photometric calibration reported here. The new $EM$ calibration for
VAULT is 1 DN s$^{-1}$ pix$^{-1}$ = $3.74\times 10^{26}$ cm$^{-5}$. To
account for integrating the energies over an area which may contain
both flaring and background (likely optically thin) emission, we have
subtracted the emission from the first, pre-event image from the
plots. The resulting energy levels are very similar to those for
coronal bright points \cite{1998ApJ...501L.213K} as the EIT
observation of plasma at $T\geq8\times 10^4$ K
suggests. Unfortunately, we cannot tell whether there is any coronal
emission from this bright point because it lies outside the
\textsl{TRACE} field of view and EIT was observing solely in \ion[He
i] during the VAULT flight. We only report counts for the EIT light
curves because there is only one wavelength available and the emission
measure cannot be calculated (right axis in Figure~\ref{fig:blob}). The VAULT and EIT curves are aligned at
the pre-event emission level along the intensity axis to allow a
comparison. The main conclusions from Figure~\ref{fig:blob} are that
the Ly$\alpha$ and \ion[He ii] have a similar impulsive phase and the
\ion[He ii] emission seems to be the extention of the cooler
Ly$\alpha$ emission. This is also in agreement with the earlier
results showing a delay from cooler to hotter coronal lines and
extends the detection of heating events to a much lower layer of the
atmosphere.
 \begin{figure}
  \centerline{\includegraphics[width=\textwidth]{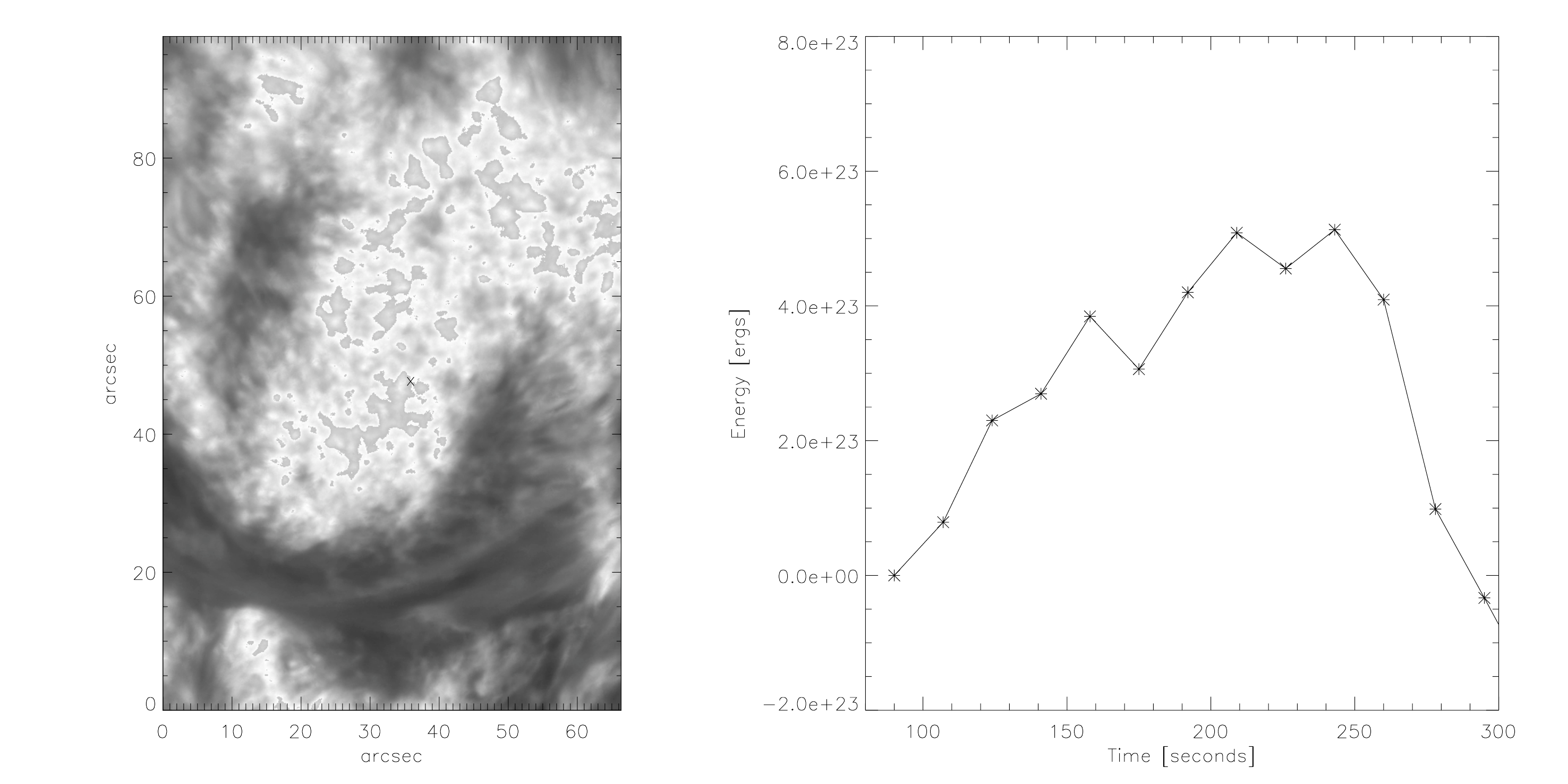}} 
  \caption{Microflaring event in the Ly$\alpha$
    plage. \textsl{Left:\/} The symbol 'x' marks the location of the
    brightening. \textsl{Right:\/} The energy estimate for this
    event.} \label{fig:blob2}
\end{figure}
The energy estimates in Figure~\ref{fig:blob} are in the range of
microflares which seems reasonable for the lower TR. An inspection of
the plage area around the filament shows fainter brightenings that
could still be classified as impulsive based on their light
curves. Energy estimates for those brightenings are around
$<5\times10^{23}$ ergs, lower than a
microflare. Figure~\ref{fig:blob2} shows an example of such a
brightening.  The energy was estimated over an area of $\sim
1.8\arcsec \times 1.8\arcsec$; all other assumptions are the same as
above. Because these brightness changes are very close to the overall
brightness variability of the plage, it is difficult to say with
certainty that these are flaring events. A more sophisticated analysis
is required but it is beyond the scope of the paper.

\section{Plage and Spicules in $Ly\alpha$\label{s:spicules}}
The active plage has been studied in some detail using the first VAULT
observations \cite{2001ApJ...563..374V}. The large degree of spatial
structuring and the variability of these structures combined with the
complex radiative character of Ly$\alpha$ emission complicate the
detailed analysis of the plage. The plage has clearly a different
morphology than the Quiet Sun. It lacks extended loop-like structures,
but contains many point-like brightenings reminiscent of the 171 \AA\
moss. Actually the \textsl{TRACE\/} 171 \AA\ images show moss over the
majority of plage with large scale loops located only in the periphery
(Figure~\ref{fig:4promin}). As expected, the moss underlies hotter
loops seen in the EIT \ion[Fe XV] 284 \AA\ images but the Ly$\alpha$
brightness is not correlated with the degree of coronal heating
above. A quick inspection of the Ly$\alpha$ and 171 \AA\ images in
Figure~\ref{fig:4promin} shows that despite the largely similar mossy
appearance, there are several areas without a detailed correlation
between corona and lower TR as noted before (e.g., region R2 in
Figure~2 of \inlinecite{2001ApJ...563..374V}). Neutral hydrogen
diffusion across field lines as proposed by
\inlinecite{2008ApJ...683L..87J} maybe an explanation of the uniform
brightness of the plage in Ly$\alpha$ but better calculations are
needed before we can establish the viability of this mechanism.

\subsection{Detection of Proper Motions}
A significant part of the variability seems quite
random. For a given pixel, the brightness change could be due to the
weakening of the emission, the lateral motion of the bright point or
the appearance of dark (likely absorbing) features. We believe that
these changes can be understood as the buffeting of the Ly$\alpha$
moss by chromospheric H$\alpha$ jets similarly to the picture proposed
by \inlinecite{1999SoPh..190..419D} for the 171 \AA\ moss but extending it
to much smaller spatial scales. 

On the other hand, we can identify coherent motions in several
places. The most obvious ones can be found at or near the filament
footpoints and along their backbone structure. Blobs of weakly
emitting Ly$\alpha$ seem to flow towards the lower atmosphere. At the
same time, apparently upward moving blobs can be seen also at the
filament footpoints as well as along the boundaries of the
small network cell within the plage and basically in most locations
where there is high contrast with the background. 
 \begin{figure}
  \centerline{\includegraphics[width=\textwidth]{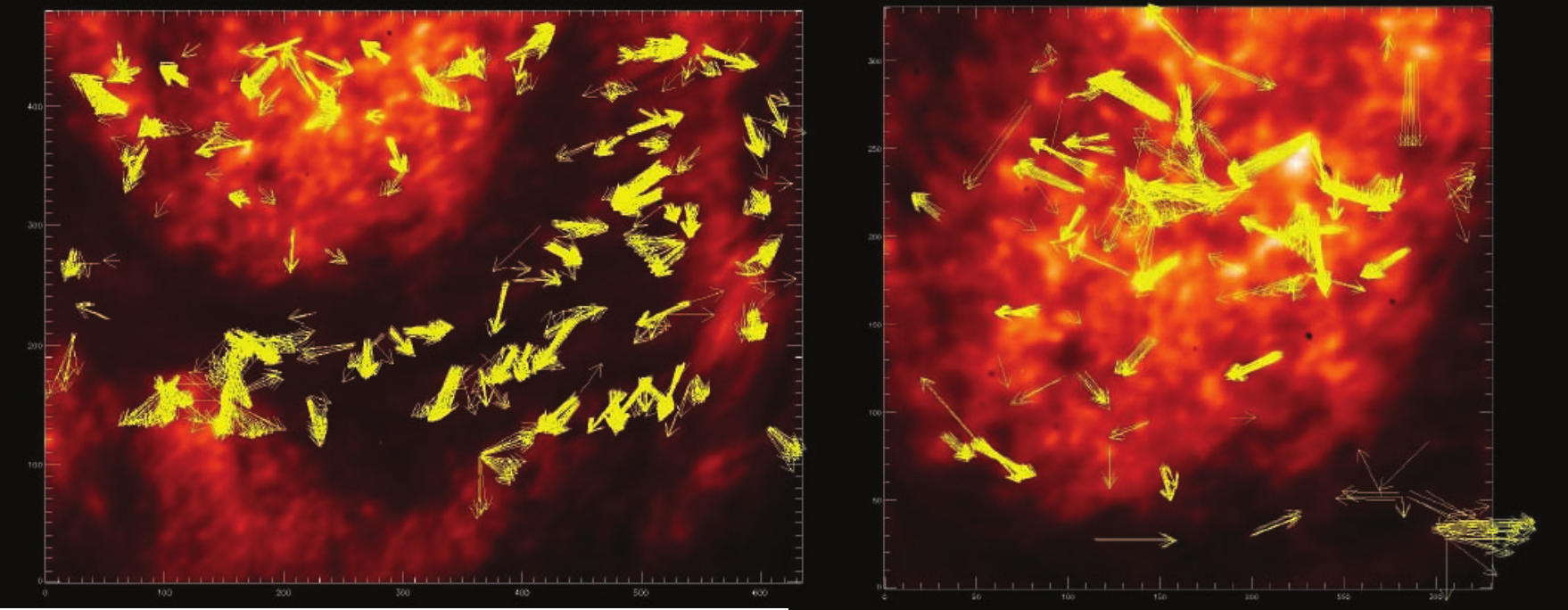}} 
  \caption{Detection of proper motions in Ly$\alpha$ plage. The
    lengths of the displacement vectors are proportional to the
    esimated speed. Only pixels with correlation coefficients $\geq
    0.3$ and intensity changes $\geq 3\sigma$ above the background are
    considered. The units are VAULT pixels
    ($0.112\arcsec$/pixel). \textsl{Left:\/} Filament and neaby
    plage. Upward motions can be seen along the western
    footpoint. \textsl{Right:\/} Plage detail. Note the
    counterstreaming motions along the filament boundary and diverging
    (explosive?) motions at certain bright
    points.} \label{fig:motions}
\end{figure}

In an attempt to quantify these motions we used a local correlation
method to track the blobs in time. To suppress the influence of the
background buffeting motions we calculated the standard deviation,
$\sigma$, of the intensity variability for each pixel at the peak of
the emission and then considered only pixels with $\geq 3\sigma$ as
inputs to the cross correlation algorithm. The large degree of
variability and spatial structuring results in many correlations. So
we kept only the pixel with correlation coefficients higher than 0.3
and estimated their speeds and velocity vectors. We derive speeds in
the range of 5-20 km s$^{-1}$ which are similar to the speeds of
chromospheric fibrils and spicules (e.g.,
\opencite{2000A&A...360..351W}; \opencite{2008ApJ...673.1194L}). In general, the
cross correlation results showed motions in all directions reinforcing
the visual impressions of the large degree of randomness in the
Ly$\alpha$ structures. However, a closer inspection of the
displacement vector revealed several instances of coherent motions. In
the example of Figure~\ref{fig:motions}, we can see upward motions
along the western filament footpoints and the filament
boundaries. There was clear evidence of counterstreaming motions along
the filament. Some of those were in the upper range of our estimated
speeds ($\sim20$ km s$^{-1}$) and are very close to H$\alpha$ measurements in
filaments \cite{1998ASPC..150...23E,2003SoPh..216..109L}.  The nearby
plage showed motions that followed the curvature of the filament
(Figure~\ref{fig:motions}, right panel). They may lie along thin, dark
strands that are part of the filament rather than the plage. Coherent
apparently upward motions were also detected at network boundaries along
spicular-like structures. The most interesting results were at
locations of diverging motions as can be seen towards the upper end of
the field (Figure~\ref{fig:motions}). Some were associated with
moderate flare-like brightenings (right panel in
Figure~\ref{fig:motions} and Figure~\ref{fig:blob2}) and may suggest
an explosive nature for these intensity changes. It is possible that
some of the TR variability seen in TR lines with coarser resolution
and attributed to stationary brightenings could actually be an effect
of spatial smoothing of the above mentioned flows. In other places, we
found diverging vectors suggesting rotation. In the wavelet-processed
movies, we see unwinding features at those areas. They are very
suggestive of the so-called mini-CMEs detected recently by
\textsl{STEREO\/} and associated with vortex flows at supergranular
boundaries \cite{2009A&A...495..319I}.

A big advantage of VAULT's large FOV is the observation of similar
structures both on disk and at the limb. An obvious candidate are the
spicules. \inlinecite{2009A&A...499..917K} measured the dynamics of
several Ly$\alpha$ spicules and found many similarities to the
H$\alpha$ dynamic fibrils despite the short VAULT time series. Based
on the \textsl{TRACE} co-alignement we measured Ly$\alpha$ spicules to
be $8\arcsec-12\arcsec$ in height, from VAULT spicule edge to the
TRACE limb position. When we consider the co-aligned cotemporal BBSO
H$\alpha$ channel, Ly$\alpha$ spicules can be up to $\sim2$\arcsec\
higher than in the comparatively optically thinner
H$\alpha$. Although scattered light may play a role in the
  ground-based observations, the height difference between Ly$\alpha$
  and H$\alpha$ appears to be significant. These results imply that
Ly$\alpha$ spicules could be the outer sheaths of the H$\alpha$ fibrils
\cite{2009A&A...499..917K}. When we take into account similar results
between H$\alpha$ and \ion[C iv] \cite{2006A&A...460..309D} it becomes
obvious that chromospheric mass is propelled to the corona via the
fibrils and undergoes heating appearing in successively higher
temperatures (Ly$\alpha$ to \ion[C iv], for example). This
  scenario seems to corroborate the very recent results of
  \inlinecite{2009ApJ...701L...1D} where it is proposed that type-II
  spicules may be the means of chromospheric plasma transport to
  coronal levels and temperatures and may play an important role in
  the coronal heating problem. Further observations of both fibrils
  and type-II spicules are, therefore, highly desirable in Ly$\alpha$
  (in addition to chromospheric and coronal lines) to provide a more
  robust connection between the evolution of the chromospheric and
  coronal structures. For the moment, the above discussion suggests
  that spicules/fibrils may provide the mass heated to
coronal temperatures (e.g., \opencite{2007PASJ...59S.655D}).

Overall, our initial attempt to characterize the variability seen in
the VAULT images seems to provide reasonable results. The most serious
problem is the large amount of variability in all intensity and
spatial levels. We plan to revisit the analysis of proper motions
using our newly available wavelet-processed images which supress the
background ``noise'' and may enhance the effectiveness of
cross-correlation techinques.

\section{Discussion \label{s:dis}}

The VAULT data, being taken from a sounding rocket platform, do not
permit long time series investigations of the Ly$\alpha$
atmosphere. However, they do provide several tantalizing clues about
the dynamics and morphology of the crucial interface of the upper
chromosphere/lower TR at least for long-lived structures and for
variability at a time scale of a few minutes.

The improved photometric analysis of the VAULT data combined with
better Ly$\alpha$ models show that, for wideband imaging at least,
most of the emission originates from the lower TR ($\geq 10^4$ K) and
only the darker areas contain much chromospheric material
(Table~\ref{tbl:meas}). Therefore, Ly$\alpha$ imaging observations are
a great probe for the structure of the transition region
\cite{2005ESASP.596E..66T}. It seems that the Ly$\alpha$ Quiet Sun is
dominated by longer thread-like structures reminiscent of H$\alpha$
fibrils. The VAULT observations have provoked new ideas about the
nature of the TR as the region where neutral hydrogen atoms from these
threads diffuse across magnetic field lines, interact with nearby
electrons and subsequently excite, ionize, and/or radiate to provide
the emission we see in TR lines \cite{2008ApJ...683L..87J}. These
ideas remain to be tested in detail but they demonstrate the value of
sounding rocket observations.

The high spatial resolution of the VAULT data resolves a great deal of
variability, mostly associated with lateral motions, in the plage. We
believe that the majority of this variability can be explained as
buffeting of the Ly$\alpha$ structures by cooler material, such as
H$\alpha$ jets. In addition, the VAULT observation of spicules show
that they extend higher and have larger widths but otherwise similar
dynamics \cite{2009A&A...499..917K} with their H$\alpha$
counterparts. These observations verify past SUMER results
\cite{1998A&A...334L..77B} and provide significant support for an
interesting idea put forth recently by
\inlinecite{2009ApJ...702.1016D} to explain the large emission measure
discrepancies between coronal and lower TR structures
\cite{2001ApJ...563..374V} as a result of EUV absorption from
chromospheric material injected in the corona. When we consider these
observations/ideas together; namely, the long network loops and
neutral cross-field diffusion, the continuous buffeting, and the
Ly$\alpha$ jets as extension of H$\alpha$ dynamic fibrils, we come to
the conclusion that the transition region may be nothing more than the
transient, evaporating part of the chromosphere rather than the stable
layer in the simple 1D models, such as \citeauthor{1981ApJS...45..635V} (1981),
long favoured in our discipline. The VAULT and more recently
\textsl{Hinode\/}/SOT observations are making us reassess our views on
the structure of the lower solar atmosphere.

The large field of view of the instrument led to observations of
basically every solar structure, with the exception of coronal holes,
which enabled us to estimate the contribution of various Ly$\alpha$
sources to the observed intensity and thereby introducing the first
empirical segmentation of Ly$\alpha$ irradiance to its sources
(Sec.~\ref{s:inten}). We found that Quiet Sun features can have intensities
several times the intensity of the average Quiet Sun and that
filaments exhibit both absorption and emission in Ly$\alpha$. The
latter can be as bright as weak bright points. Optically thin
structures, up to 50\% fainter than the average Quiet Sun may exist in the
center of cell interiors and as off limb loops. We did find that high
temperatures are likely in off-limb Ly$\alpha$ loops which may explain
their large heights ($\sim60\arcs$, corresponding to $\approx 45,000$~km) in the
VAULT images. We also found that active region filament partially absorbs plage
emission, by around 20\% to 30\%, and this effect may need
to be considered carefully in irradiance studies. These segmentation
results may be useful to irradiance studies until a full disk
Ly$\alpha$ imaging becomes available.

The VAULT images provide the first ever unambiguous Ly$\alpha$ imaging
of the fine structure of filaments/prominences and show that both
emission and absorption takes places along the prominence backbone. It
is interesting to note, that the underlying plage is visible through
several locations along the prominence suggesting that Ly$\alpha$ is
optically thin and that the distribution of hydrogen is highly
anisotropic through these structures. It is also clear that the
Ly$\alpha$ filament is larger than the H$\alpha$ one
\cite{Millard:2009jk} and is likely to reach a higher altitude. The
high LRI measurements in the filaments (up to 5, Table~\ref{tbl:meas})
are again consistent with a decreased optical thickness, even to the
point of being optically thin. According to \inlinecite{1986A&A...154..154G}
and their Figure~5, a very hot temperature of $\sim5\times 10^4$K is
also possible. For this study we have chosen the cooler more plausible
solution of the curve. Nevertheless, with the lack of other
observational constraints, it remains unsolved whether the hot
solution is possible. One approach would be a point-to-point
correlation with other chromospheric-TR lines. This calls for a
high-resolution spectrograph which is not currently available in
space.

VAULT images also reveal a wealth of activity in both the
plage and the Quiet Sun regions. In the latter, we see evidence of
braiding in the loop structures outlining the cell
boundaries. However, we do not see any direct unambiguous evidence of
reconnection as would be expected from such activity. It may be that
longer time series are needed to evidence such events. Alternatively,
a mixture of cool absorbing structures propagating alongside these
loops may create the appearance of braiding. Those structures may be
the same absorbing structures that create the buffeting motions in the
plage. On the other hand, we
see frequent brightenings and even jets in the interior of the
cells. This is the first time that the Ly$\alpha$ emission from these
areas has been imaged and the amount of observed activity was
unexpected. The brightenings seem to be associated with the emergence
of magnetic field elements and their subsequent movement towards the
cell boundary. These motions are regularly seen with sub-arcsecond
resolution magnetographs (e.g., SOUP instrument,
\opencite{1989ApJ...336..475T}) but we did not have any available
during the flight. The relation between the emerging flux and the
Ly$\alpha$ brigtenings remains to be confirmed in a future flight but
if it is true it suggests that the effects of even such small magnetic
elements reach substantial heights in the solar atmosphere. We wonder
whether some of those jets  are the Ly$\alpha$ counterparts of the Type II
spicules seen in the SOT observations \cite{2007PASJ...59S.655D}.

Another rather surprising observation is the relative scarcity of
microflaring events. We have been able to identify a handful in the five
minutes of observation. These have been previously
identified by \inlinecite{1998A&A...336.1039B}. They suggest a
possible link with atmospheric turbulence. In our observations we have
observed them in both active regions and Quiet Sun regions. The largest
of them had a light curve and energy consistent with a microflare and
was detected in \ion[He ii] as well (Figure~\ref{fig:blob}). Others had
energies in the range of $10^{24}$ ergs. Although the
short duration of the observation does not allow proper statistics for
the occurrence of these events, our field of view covers a substantial
part of the solar disk. Therefore, it seems unlikely that
microflares are a common occurrence in this temperature range.

\section{Conclusions \label{s:con}}
We conclude our overview of Ly$\alpha$ imaging observations with a set
of ``lessons learned'' that may be useful in the design of future
Ly$\alpha$ instruments or observation campaigns.
\begin{itemize}
\item Ly$\alpha$ is formed at the critical interface between the upper
  chormosphere and the low TR. Thus, imaging is very useful and the
  well-known difficulties surrounding the interpretation of Ly$\alpha$
  emission are not insurmountable anymore.  We can rely on models to
  derive reasonable physical parameters for the observed structures.
\item We see few Ly$\alpha$ structures close to the instrument
    resolution limit of $0.5\arcsec$. Only absorbing (dark) features and
    off-limb structures (in emission) can be identified at that
    resolution. Most of the on-disk structures are much larger. This
    could be due to the high optical thickness of the line throughout
    these structures. In any case, this observation should be
    considered in the design of future Ly$\alpha$ telescopes. Extreme
    resolutions may not be useful unless the instruments can
    spectrally resolve the line or their science objectives include
    absorption or off-limb features.
\item There is evidence of optically thin emission in many locations
  besides the obvious limb structures. Areas around filaments are
  especially interesting.  This would require observations of the
  spectral profiles to be confirmed.
\item There is considerable structure and variability within the cell
  interiors which is probably linked to photospheric flux
  emergence. This is a new area for Ly$\alpha$ studies and to
  understand it will require a future telescope sensitivity equal or
  better than VAULT. 
\item Even if flaring activity is relatively unimportant,
  there is variability. Future instruments should achieve
  both high signal-to-noise ratio and cadence to allow the study of
  both.
\item Both types of spicules are observable and given the significant temperature
  range of Ly$\alpha$ formation, observations in Ly$\alpha$ are
  excellent tracers of the injection of material from the chomosphere
  to the oorona.
\end{itemize}  

For the near future, the advent of \textsl{Hinode\/}/SOT has created a
new and unique opportunity to address the nature of the transition
region by combining VAULT and SOT observations of Quiet Sun structures
and spicules. We plan to seek funding for refurbishment of the VAULT
payload, which was damaged during its last flight, and for an
underflight with SOT with the specific objectives of addressing the
nature of the long Ly$\alpha$ fibrils over the quiet network and
investigate type-II spicule dynamics particularly at coronal
holes. However the dynamics of the type-I spicules and macro-spicules
may need longer time series due to their longer lifetimes \cite{2005A&A...438.1115X}. 
 
To summarize, we presented a broad overview of the morphology and
dynamics of the Sun's Ly$\alpha$ atmosphere; an important but rarely
imaged region. These were the first sub-arcsecond, high sensitivity
observations of this line and, at that time, the highest resolution
observations of any solar structure from space. The VAULT observations
showed that Ly$\alpha$ emission arises from every location and in
every solar feature, and generated new ideas about the nature of the
transition region and coronal heating. These results demonstrate the
wideranging value of sounding rocket experiments despite their short
observing windows.

 \begin{acks}
   This work is dedicated to the memories of D. Prinz, G. Bruecker,
   and D. Lilley whose efforts have contributed enormously to the
   success of the NRL sounding rocket programs. We are grateful to
   V. Yurchyshyn for providing calibrated BBSO H$\alpha$ images, and
   to J. Cook, J. Koza, S. Martin, R. Rutten, J.C. Vial, for useful
   discussions and constant encouragement. The achievements presented
   in this paper are the product of many years of development work at
   the Naval Research Laboratory Solar Physics Branch and the NASA
   sounding rocket program.  The VAULT instrument borrows heavily from
   the High Resolution Telescope and Spectrograph. The NRL rocket team
   of J. Smith, R. Moye, R. Hagood, R. Feldman, J. Moser, D. Roberts,
   T. Spears and R. Waymire did a superb job in preparing and
   launching the instrument. We would like to acknowledge the efforts
   of the sounding rocket support team that made the VAULT launches
   possible.  We would like to particularly acknowledge the following
   individuals.  Tracy Gibb did a superb job managing the launch of
   the VAULT payload.  Frank Lau managed the development of the Mark 7
   digital SPARCS attitude control system.  We would like to
   acknowledge the SPARCS team for their superb efforts in the support
   of our launch.  Jesus and Carlos Martinez developed and operated
   the Mark 7 SPARCS.  Richard Garcia, Shelby Elborn and Kenneth Starr
   developed the VAULT telemetry section.  The support from the White
   Sands Missile Range and Wallops Flight Facility NASROC personnel
   was of the highest caliber.  The VAULT instrument development work
   has been supported by the ONR task area SP033-02-43 and by NASA
   defense procurement request S-84002F.
 
\end{acks}

\bibliographystyle{spr-mp-sola-cnd}
\bibliography{vault-biblio}  

\end{article} 
\end{document}